\documentclass[sigconf]{acmart}

\usepackage{natbib}
\AtBeginDocument{%
  }

\usepackage{geometry,spath3,tikz}
\usepackage{tikz-cd}
\usetikzlibrary {shapes.multipart, positioning, arrows, calc, decorations.pathreplacing}
\usetikzlibrary{arrows.meta, fit, shapes}
\usetikzlibrary{calligraphy}
\usepackage{amsmath, pifont}
\usepackage{mathpartir, hyperref}
\usepackage{amsfonts}
\usepackage{amsthm}
\usepackage{pl-syntax}
\usetikzlibrary{shapes}
\usepackage{float}
\usepackage[normalem]{ulem}
\usepackage{stmaryrd}
\usepackage[linesnumbered,ruled,vlined]{algorithm2e}
\usepackage{algpseudocode}
\usepackage{listings}
\usepackage{nameref, xparse}
\usepackage{pgfplotstable}
\usepackage{pgfplots}
\usepackage{siunitx}
\usepackage{caption}
\usepackage{subcaption}
\usepackage{wrapfig}
\usepackage{xspace}
\usepackage{graphicx}
\usepackage[inline]{enumitem}
\usepackage{multirow}
\usepackage{causets}
\usepackage{arydshln}
\usepackage{tabularx}
\usepackage{hhline}
\usepackage{colortbl}
\usepackage{arydshln}
\usepackage{setspace}
\newcommand{\Si}{\Sigma}

\newcommand{\om}{\omega}
\newcommand{\Ga}{\Gamma}
\newcommand{\De}{\Delta}
\newcommand{\eps}{\epsilon}

\newcommand{\integer}{\code{int }}

\newcommand{\lwe}{\code{LWE}\xspace}
\newcommand{\rlwe}{\code{RLWE}\xspace}
\newcommand{\gsw}{\code{RGSW}\xspace}
\newcommand{\glwe}{\code{GLWE}\xspace}
\newcommand{\id}{\code{id}}

\newcommand{\tht}{\code{ct}}
\newcommand{\low}{\inf}
\newcommand{\high}{\sup}

\newcommand{\rulename}[1]{\textsc{#1}}
\newcommand{\mb}{\mathbb{Z}}

\NewDocumentCommand{\plr}{m m}{\ensuremath{\code{plain \,}\langle #1, #2 \rangle}}
\NewDocumentCommand{\cimod}{m m m m}{\ensuremath{\code{cipher \,} \langle #3, #4, #1, #2 \rangle}}

\NewDocumentCommand{\ring}{m}{R_#1}
\NewDocumentCommand{\cring}{m m}{R^{#2}_{#1}}

\NewDocumentCommand{\jdg}{m m m}{ \pparam;#1 \vdash #2:#3 }

\NewDocumentCommand{\jdgt}{m m m O{2}}{ #1^2 \vdash_{#4} #2:#3 }
\NewDocumentCommand{\jdgsi}{m m m m O{\De}}{ #1, #5 \vdash_{#4} #2:#3 }
\NewDocumentCommand{\relaxjdg}{m m m O{\Si}}{ #1, \De \vdash^r_{#4} #2:#3 }

\NewDocumentCommand{\lang}{m O{}}{\ensuremath{\code{ILA}_{#1}^{#2}}}
\NewDocumentCommand{\ila}{}{ILA\xspace}
\NewDocumentCommand{\ilabgv}{}{\ensuremath{\text{ILA}_{{(\it bgv) }}}\xspace}
\NewDocumentCommand{\ilatfhe}{}{\ensuremath{\text{ILA}_{{(\it tfhe) }}}\xspace}
\NewDocumentCommand{\ilabfv}{}{\ensuremath{\text{ILA}_{{(\it bfv) }}}\xspace}
\NewDocumentCommand{\modswitch}{m}{\ensuremath{\code{modswitch}(#1)}}
\NewDocumentCommand{\code}{m}{\ensuremath{\mathsf{#1}}}

\NewDocumentCommand{\vecs}{m}{\mathsf{vec}\, #1}

\newcommand{\Sort}{\mathcal{S}}
\newcommand{\sort}{\mathsf{sort}}
\newcommand{\PSort}{\mathsf{plain}}
\newcommand{\CSort}{\mathsf{cipher}}
\newcommand{\MSort}{\mathsf{msg}}

\newcommand{\Cipher}{\mathsf{cipher}}
\newcommand{\Plain}{\mathsf{plain}}
\newcommand{\Msg}{\mathsf{msg}}
\newcommand{\interp}{\mathsf{interp}}

\newcommand{\eif}[3]{\mathsf{if}\ #1\ \mathsf{then}\ #2\ \mathsf{else}\ #3}

\newcommand{\op}{\mathsf{op}}
\newcommand{\bnd}{\mathsf{bnd}}
\newcommand{\topub}{\mathsf{topub}}
\newcommand{\msg}{\mathsf{msg}}
\newcommand{\sem}[1]{\llbracket #1 \rrbracket}
\newcommand{\cskip}{\mathsf{skip}}
\newcommand{\config}[1]{\langle #1 \rangle}
\newcommand{\ppset}{\mathcal{PP}}
\newcommand{\spset}{\mathcal{SP}}
\newcommand{\pparam}{\mathsf{pp}}
\newcommand{\sparam}{\mathsf{sp}}
\newcommand{\judgment}[1]{\fbox{\ensuremath{#1}}}

\newtheorem{theorem}{Theorem}[section]

\newtheorem{definition}{Definition}[section]

\makeatletter
\providecommand*{\Dashv}{%
  \mathrel{%
    \mathpalette\@Dashv\vDash
  }%
}
\newcommand*{\@Dashv}[2]{%
  \reflectbox{$\m@th#1#2$}%
}
\makeatother

\DeclareCaptionLabelFormat{mywraplabel}{\hspace{.5ex}#1~#2.\newline}
\definecolor{codegreen}{rgb}{0,0.6,0}
\definecolor{codegray}{rgb}{0.5,0.5,0.5}
\definecolor{codepurple}{rgb}{0.58,0,0.82}
\definecolor{backcolour}{rgb}{0.95,0.95,0.92}
\lstdefinestyle{mystyle}{
    backgroundcolor=\color{backcolour},   
    commentstyle=\color{codegreen},
    keywordstyle=\color{magenta},
    numberstyle=\tiny\color{codegray},
    stringstyle=\color{codepurple},
    basicstyle=\footnotesize,
    breakatwhitespace=false,         
    breaklines=true,                 
    captionpos=b,                    
    keepspaces=true,                 
    numbers=left,                    
    numbersep=5pt,                  
    showspaces=false,                
    showstringspaces=false,
    showtabs=false,                  
    tabsize=2
}

\newcommand{\capfig}[2]{\textbf{#1}. {\textit{#2}}}

\begin{document}

%\title{ILA: A Formal Model for Sound Noise Management in Fully Homomorphic Encryption Schemes}
 \title{ILA: Correctness via Type Checking for Fully Homomorphic Encryption} 

\author{Tarakaram Gollamudi}
\affiliation{%
  \institution{Independent Researcher}
  \country{USA}}
\email{gollamudi.ram@gmail.com}

\author{Anitha Gollamudi}
\affiliation{%
  \institution{University of Massachusetts Lowell}
  \city{Lowell}
  \state{MA}
  \country{USA}}
\email{anitha_gollamudi@uml.edu}

\author{Joshua Gancher}
\affiliation{%
  \institution{Northeastern University}
  \city{Boston}
  \state{MA}
  \country{USA}}
\email{jgancher@northeastern.edu}

\renewcommand{\shortauthors}{Tarakaram Gollamudi, Anitha Gollamudi, and Joshua Gancher}

\begin{abstract}
RLWE-based Fully Homomorphic Encryption (FHE) schemes add some small
    \emph{noise} to the message during encryption. The noise accumulates with
    each homomorphic operation. When the noise exceeds a critical value, the FHE
    circuit produces an incorrect output. This makes developing FHE applications
    quite subtle, as one must closely track the noise to ensure
    correctness. However, existing libraries and compilers offer
    limited support to statically track the noise. Additionally, FHE circuits
    are also plagued by wraparound errors that are common in finite modulus
    arithmetic. These two limitations of existing compilers and libraries 
    make FHE applications too difficult to develop with confidence.  

In this work, we present a \emph{correctness-oriented} IR, Intermediate Language for Arithmetic circuits, for type-checking circuits intended for homomorphic evaluation. Our IR is backed by a
type system that tracks low-level quantitative bounds (e.g., ciphertext noise)
without using the secret key.
Using our type system, we identify and prove a strong \emph{functional
correctness} criterion for \ila circuits.
Additionally, we have designed \ila to be maximally general: our core type system does not
directly assume a particular FHE scheme, but instead axiomatizes a \emph{model}
of FHE. We instantiate this model with the exact FHE schemes (BGV, BFV and TFHE), and obtain functional correctness for free.

We implement a concrete type checker \ila, parameterized by the noise estimators
for three popular FHE libraries (OpenFHE, SEAL and TFHE-rs).
Additionally, our novel algorithm for inferring the placement of \emph{modulus switching}, a common noise management operation, uses the \ila type checker as a validation framework to ensure the correctness of the optimization. Evaluation shows that \ila type checker is sound (always detects noise overflows), practical (noise estimates are tight) and efficient.
  
\end{abstract}

%% %%
%% %% The code below is generated by the tool at http://dl.acm.org/ccs.cfm.
%% %% Please copy and paste the code instead of the example below.
%% %%
%% \begin{CCSXML}

%% \end{CCSXML}

%% \ccsdesc[500]{Do Not Use This Code~Generate the Correct Terms for Your Paper}
%% \ccsdesc[300]{Do Not Use This Code~Generate the Correct Terms for Your Paper}
%% \ccsdesc{Do Not Use This Code~Generate the Correct Terms for Your Paper}
%% \ccsdesc[100]{Do Not Use This Code~Generate the Correct Terms for Your Paper}


\maketitle

\section{Introduction}
Unlike traditional encryption schemes, a \emph{fully homomorphic encryption}
(FHE) enables computations on encrypted data. Popular FHE schemes---such as
BGV~\cite{BGV2012}, CKKS~\cite{ckks}, TFHE~\cite{tfhe}---that are based on Ring
Learning With Errors (RLWE) adds some small \emph{noise} to the message to make
the encryption non-deterministic (and thus secure). This noise accumulates with each homomorphic operation. A crucial invariant that needs to be preserved
for the successful decryption is that the noise must not exceed a \emph{critical
value}; otherwise, the decryption will not represent the accurate value.
Moreover, noise overflows are attack vectors for secret key recovery
attacks~\cite{ckks_security}. Noise management operations such as \emph{modulus
switching} and \emph{programmable bootstrapping} (PBS) reduce noise but are
computationally expensive, requiring judicious use. Additionally, all these
schemes operate in finite rings (e.g., $\frac{Z_q[x]}{x^d-1}$); a value overflow could potentially corrupt the output.
Thus, FHE application developers need tools that guarantee correct evaluation.

Unfortunately, none of the existing FHE  compilers \cite{eva,chielle2018e3,
fhelipe, alchemy} or libraries---such as OpenFHE~\cite{openfhe},
SEAL~\cite{sealcrypto}, HELib~\cite{helib}, TFHE-rs~\cite{tfhe-rs},
Lattigo~\cite{lattigo}--- guarantee correct FHE evaluation in the presence of
overflows in noise or value.
Indeed, to avoid incorrect outputs, 
they either rely on programmer expertise or require one to simulate many runs of
the FHE circuit, which 
carries a high
computational cost and thus harms iterative development.
Additionally, \emph{no} realistic FHE library to date can simultaneously reason
about overflows in both RLWE-induced noise and overflows in the underlying
encrypted values. 

Thus, none of the existing libraries offer a generic yet robust correctness reasoning framework that statically and automatically detects overflows. 

In response, we present \ila, a \emph{correctness-oriented} intermediate representation (IR) for \emph{statically typechecking} a \emph{family} of RLWE-based circuits intended for homomorphic evaluation. \ila is backed by a type system that tracks low-level quantitative bounds (e.g., ciphertext noise, value ranges) without using the secret key.
Using our type system, we identify and prove a strong \emph{functional
correctness} criterion for \ila circuits: if $C$ is well-typed, then
homomorphically evaluating $C$, and then decrypting, is equivalent to 
running a cleartext version $C_\mathsf{clear}$ on its decrypted inputs. 
In essence, this correctness criterion guarantees that FHE does not introduce
any additional functional bugs, relative to the original, non-homomorphic
circuit. 

A crucial insight for \ila is that RLWE-based FHE schemes share the
same high-level characteristics for homomorphic and noise management operations.
Thus, we have designed \ila to be maximally general: our core type system does not
directly assume a particular FHE scheme, but instead axiomatizes a \emph{model}
of FHE.  Thus, \ila represents a family of FHE models.
In turn, we are able to instantiate this model 
with the widely used non-approximate schemes: BGV, BFV and TFHE.

Importantly, all \ila model instantiations inherit functional correctness for free.
This is notable as the correctness guarantees can be used to validate noise-preserving transformations. For instance, \ila's BGV instantiation validates an optimization involving modulus switching. Similarly, \ila's TFHE instantiation decouples standard homomorphic and PBS operations; this enables deploying faster but correct TFHE circuits that are PBS-free.

We do not intend \ila to be used directly, but instead to provide a layer of
correctness validation on top of an existing pipeline for FHE. To demonstrate
this use case, we implement an extensible \ila compiler with static type checkers parameterized by the scheme. \ila compiler supports multiple FHE backends (e.g., SEAL, OpenFHE and TFHE-rs) for BGV, BFV and TFHE schemes. Using the type checker, one can obtain, for free, an FHE circuit that is formally verified for functional correctness.

Our evaluation shows that the overflow analysis employed in the type checker is sound (no false negatives) and practical (fewer false positives). More importantly, the entire analysis is static: neither a secret key nor circuit execution is required. This brings down the application debugging costs by enabling offline development.

Finally, we present a novel insight on the placement of \emph{modulus switching} (MS),  a common noise managing optimization for FHE circuits. Developers use MS to improve the multiplicative depth (the longest sequence of ciphertext multiplication) of the circuit. However, one cannot simply annotate the entire circuit with MS operations as they are expensive. Moreover, incorrect placement could counteract any potential multiplicative depth gains; worse the circuit may even fail to run as modulus-switched ciphertexts can only operate with equivalently switched ciphertexts.

Our MS inference procedure is guided by the type checker: it not only picks placement that maximizes the remaining noise budget but also validates the correctness of the resulting circuit.

\noindent
\textbf{Contributions.} In summary, we:
\begin{itemize}
\item present \ila, a generic and formal intermediate language and type system that embeds the notion of overflow(s) for FHE circuits in a scheme-agnostic manner;
 \item formally specify and prove a (functional) correctness theorem for \ila, which guarantees that well-typed  circuits evaluate correctly;
\item instantiate \ila with BGV, BFV and TFHE, the three  popular absolute schemes, demonstrating the generality of \ila. Notably, TFHE  operations are decoupled from the expensive programmable bootstrapping enabling users to deploy not only functionally correct but also runtime-efficient FHE circuits;
 \item transformation validation of our novel  modulus switching inference algorithm;
\item implement \ila compiler targeting three backends and a static type checker that detects both noise  and value overflows for all the three schemes; importantly, the type checker does not require a secret key to estimate the noise in the ciphertext nor does it run the expensive FHE circuit; and
 \item demonstrates that \ila is both expressive and effective with tight noise bounds; \ila's static approach at detecting overflows outperforms the state-of-the-art dynamic approach by orders of magnitude.
\end{itemize}

We make our implementation and evaluation case studies available as an anonymous GitHub repository\footnote{\url{https://github.com/anon-ila/ila}}.

\vspace{-1em}
\section{Background}\label{sec:background}
A \textbf{R}ing \textbf{L}earning \textbf{W}ith \textbf{E}rrors-based encryption scheme contains three steps: First, a probabilistic keygen algorithm that on input $\lambda$, generates a secret key $sk$ and a public key $pk$. Second, a probabilistic encryption algorithm generates a ciphertext for a given $pk$ and a message $m$ taken from the message space ${\MSort}$. The encryption can be further seen as a two-step process, the first step encodes a message $m$ to an element $p$ of the \emph{plaintext} space. The second step encrypts the plaintext $p$ and adds a random noise for security generating an element $\tht$ of the \emph{ciphertext} space. Finally, a deterministic decryption algorithm takes a ciphertext $\tht$ and $sk$ and emits a plaintext message $p$ followed by a decoding step that emits the corresponding message $m$. 

In an RLWE-based scheme, the plaintext space is a ring. The plaintext space encodes a message (or cleartext) space. The message space is usually $\mathbb{Z}_t$ (the ring of integers modulo $t$) or $\mathbb{Z}_t^d$ ($d$ tuples of $\mathbb{Z}_t$). The plaintext space is usually the ring $\ring{t} = \frac{\mathbb{Z}_t[x]}{(x^d+1)}$. The \emph{plaintext modulus} $t$ is a positive integer and $d$ is usually a power of $2$. Elements of this ring may be seen as polynomials of degree up to $d-1$ with coefficients in $\mathbb{Z}_t$. Addition and multiplication are usual polynomials addition and multiplication modulo $(t, (x^d+1))$. The ciphertext space is $\cring{q}{n+1}$, where the \emph{coefficient modulus} $q$ is a large positive integer. Elements of this space are $n+1$ tuples with each component a polynomial from $\ring{q}$. For simplicity, the reader can assume $n=1$.

\begin{figure}[t]
\begin{tikzpicture}[every fit/.style={inner sep=0pt, outer sep=0pt, draw}]
\begin{scope}[y=0.5cm]
\node (A) [fit={(0,0) (1,1)},label={[anchor=north west]south west:q}, fill=green!100] {};
\node  [fit={(1,0) (2.85,1)}, fill=orange!90] {$f(\eps_1, \eps_2)$};
\node [fit={(2.85,0) (3.15,1)},label={[anchor=north west]south west:t}, fill=yellow!100] {};
\node [fit={(3.15,0) (4,1)}, fill=blue!30] {$m_1 m_2$};
\end{scope}

\begin{scope}[yshift=1.2cm, y=0.5cm]
\node [fit={(0,0) (2.5,1)},label={[anchor=north west]south west:q},fill=green!100] {};
\node (B) [fit={(2.5,0) (2.85,1)}, fill=orange!90] {$\eps_2$};
\node [fit={(2.85,0) (3.5,1)}, ,label={[anchor=north west]south west:t},fill=yellow!100] {};
\node [fit={(3.5,0) (4,1)}, fill=blue!30] {$m_2$};
\end{scope}

\begin{scope}
\draw[-{Stealth[scale=1]}, line width=1pt,   anchor=north, double]
  (B.south) -- ++(0, -0.7cm);
\end{scope}

\begin{scope}[yshift=2cm, xshift=1.5cm]
  \node  {\Huge $\otimes$};
\end{scope}

\begin{scope}[yshift=2.3cm, y=0.5cm]
\node [fit={(0,0) (2.5,1)},label={[anchor=north west]south west:q},label={[anchor=south west]north west:MSB}, fill=green!100] {};
\node [fit={(2.5,0) (2.85,1)}, fill=orange!90] {$\eps_1$};
\node [fit={(2.85,0) (3.5,1)}, label={[anchor=north west]south west:t},fill=yellow!100] {};
\node [fit={(3.5,0) (4,1)}, fill=blue!30] {$m_1$};
\end{scope}

\begin{scope}[xshift=4.5cm, y=0.5cm]
\node [fit={(1.5,0) (2.65,1)},label={[anchor=north west]south west:q'}, fill=green!100] {};
\node [fit={(2.65,0) (3,1)}, fill=orange!90] {$\eps'$};
\node [fit={(3,0) (3.5,1)},label={[anchor=north west]south west:t}, fill=yellow!100] {};
\node [fit={(3.5,0) (4,1)}, fill=blue!30] {$m$};
\end{scope}

\begin{scope}[xshift=4.5cm, yshift=1.5cm, y=0.5cm]
\node [fit={(0,0) (2,1)},label={[anchor=north west]south west:q}, fill=green!100] {};
\node (C) [fit={(2,0) (3,1)}, fill=orange!90] {$\eps$};
\node [fit={(3,0) (3.5,1)}, label={[anchor=north west]south west:t},fill=yellow!100] {};
\node [fit={(3.5,0) (4,1)}, fill=blue!30] {$m$};
\end{scope}

\begin{scope}
\draw[-{Stealth[scale=1]}, line width=1pt,   anchor=north, double]
  (C.south) -- ++(0, -1cm) node [left, midway] {\code{modswitch}};
\end{scope}

\end{tikzpicture}

\begin{tikzpicture}[
  queue/.style={rectangle split, rectangle split horizontal,rectangle split draw splits = false,rectangle split part fill = {#1} , rectangle split parts=1, draw, anchor=center, minimum height = 0.3cm}
  ]
\node[minimum width=0.3cm, queue = {green!100}, label= east:noise budget](leg1) {\nodepart[text width=0.3cm, align=center]{one}};
 \node[minimum width=0.3cm, queue = {orange!90}, label= east:noise](leg2) [right= 1.9cm of leg1]{\nodepart[text width=0.3cm, align=center]{one}};
 \node[minimum width=0.3cm, queue = {yellow!100}, label= east:plaintext budget](leg3) [right= 0.9 cm of leg2]{\nodepart[text width=0.3cm, align=center]{one}};
 \node[minimum width=0.3cm, queue = {blue!30}, label= east:message](leg4)[right= 2.25cm of leg3] {\nodepart[text width=0.3cm, align=center]{one}};
 %\draw [<-, thick] (queue1.two) -- (noise) ;
\end{tikzpicture}
\caption{Noise changes during binary and unary homomorphic operations. The noise grows multiplicatively with ciphertext multiplication (left). Modulus switching reduces both noise and the remaining noise budget (right). \label{fig:hom-ops}}
\end{figure}
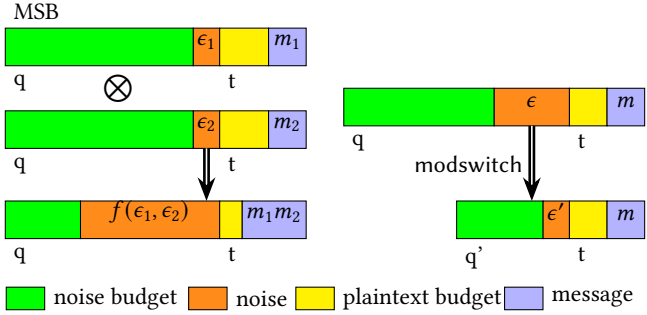

\paragraph{Absolute RLWE FHE Schemes.} BGV, BFV and TFHE are absolute RLWE FHE schemes: the decrypted output of an encrypted computation is the same as the computation on the underlying message. By contrast, CKKS is an approximate scheme: the decrypted output contains a small error. In this work, we focus on absolute schemes. All of them operate on integers and support (binary) arithmetic homomorphic operations, $\otimes$ and $\oplus$ for homomorphic multiplication and addition, respectively, and $\times$ for scalar multiplication of a ciphertext.   BGV  also supports \code{modswitch}, a modulus switching (unary) operation that reduces the noise in a ciphertext.
TFHE supports three different types of ciphertexts---\lwe, \rlwe and \gsw---and different multiplication operators---$\otimes$, $\boxtimes$, and $\boxdot$---indexed by the ciphertext type. This makes the scheme more challenging.

\paragraph{Noise changes} Homomorphic operations change the ciphertext noise. Figure \ref{fig:hom-ops} illustrates the noise changes for multiplication and modulus switching. A ciphertext is represented as a polynomial. The length (number of bits) of each coefficient is given by coefficient modulus $q$. The message $m_1$ can grow until the size of the plaintext modulus is given by $t$ bits. Every ciphertext has a current noise $\eps_1$ that can grow until $(q-t)$ bits. If it grows beyond, then an overflow occurs.

In a ciphertext multiplication, the noise growth is proportional to the product of noises in the individual ciphertexts. Modulus switching refreshes the noise---it reduces the size of the coefficient modulus by switching to a different level. In doing so, it reduces both the noise and noise budget leaving the message size and plaintext modulus unchanged. However, it reduces noise more than the noise budget. This relative gain in the noise budget enables more homomorphic operations.

\paragraph{Programmable Bootstrapping (PBS)}
Like modulus switching, PBS is also a noise management operation, however, it not only refreshes more noise budget but is also computationally more expensive. TFHE uses PBS to apply a lookup table (LUT) to LWE ciphertext(s). Technically, it is a composition of modulus switching, blind rotation and sample extraction operations. For PBS to succeed, there should be a non-zero noise budget. Since multiplication is expensive, TFHE tightly couples it with PBS. For other operations, PBS is added after a fixed number of operations.

%\vspace{-1em}
\section{Overview}\label{sec:overview}
%We introduce the key concepts of \ila  informally.
\ila is a family of intermediate languages for developing FHE applications. At a high level, \ila is a standard imperative language with abstract homomorphic operations and types parameterized by the FHE scheme. Our key insight is driven by the observation that all RLWE-based FHE schemes share common characteristics such as the scheme parameters selection, cipher and plaintext construction as well as the propagation of noise through the circuit. As a part of parameter selection, all schemes choose a set of keys (public, private, and evaluation), coefficient modulus $q$, plaintext modulus $t$ and a poly modulus degree $d$. Ciphertext and plaintexts are constructed as elements of finite rings involving the above chosen parameters. 

The core language provides a common semantic framework for statically reasoning about the correctness of  RLWE-based FHE schemes.  Each scheme instantiates \ila with scheme-specific operations. For example, arithmetic operations on ciphertexts are common in all schemes, whereas noise management operations are only selectively supported. Both BGV and TFHE support modulus switching but only TFHE supports PBS. As described in Section~\ref{sec:background}, TFHE also supports multiple operations on different types of ciphertexts having distinct properties.

For an instantiation to be valid, it has to meet certain requirements (described
in Section~\ref{sec:ila}). A valid instantiation inherits, for free, all the correctness guaranteed offered by \ila. In Section~\ref{sec:ilamodels}, we describe \ilabgv, \ilabfv and \ilatfhe, instantiations for BGV, BFV and TFHE schemes, respectively. Furthermore, we prove that they are valid.

%% \ilabgv instantiates \ila with the BGV scheme-specific operations. Importantly, it supports homomorphic addition, multiplication and modulus switching. To reason about their correctness, \ilabgv's type system tracks cryptographic details such as moduli for coefficient rings, levels for modulus switching, and error bounds induced from RLWE. Given a well-typed program, \ilabgv can run the arithmetic circuit against an existing FHE library with correct evaluation. For our implementation, we use SEAL~\cite{sealcrypto}.

\begin{figure}
\begin{lstlisting}[language=Python, style=mystyle, firstline=5, lastline=23, mathescape=true]
# Private Set Intersection between encrypted sets A and B
result = 1
n_unit = -1

for j = len$_A$ downto 0: # begin outer loop
    t1 = B[j]           # index into set B
    t2 = n_unit $\otimes$ t1    # compute -B[j]
    t3 = random_data[j]
    t4 = result $\otimes$ t3
    i = len$_B$
    while i = len$_B$ downto 0: # begin inner loop
        t5 = A[i]        # index into set A
        t6 = t5 $\oplus$ t2     # t6 = A[i] - B[j]
        t7 = result $\otimes$ t6
        result = t7 $\otimes$ t4 # 0 if PSI is non-empty


\end{lstlisting}
\caption{\capfig{\ila program for private set intersection (PSI) over encrypted
    sets A and B.}{The sets are represented as vectors; operators $\oplus$ and
    $\otimes$ represent homomorphic addition and multiplication.}
    \label{fig:psi}}
\end{figure}

Figure~\ref{fig:psi} shows the implementation of private set intersection (PSI)  in \ila~\cite{cryptoeprint:2017/299}. The goal of PSI is to find if the intersection between  two encrypted sets, A and B, (owned by different parties but encrypted using the public key owned by one of the parties) is non-empty. This is computed by subtracting each element of A from each element of B, and  accumulating the difference in result. However, a naive implementation will be insecure as the decrypted result might leak more information about the other set. In the secure version, the computations are mixed with randomized data.

In the program, sets A and B are represented as vectors; \code{A[i]} denotes the element at index $i$.  Both A and B are initialized encrypted vectors (e.g., \code{A = cipher\_init[1, 3, 36]}).  Elements of A and B are iterated using \code{while} loops and the subtraction is computed using the combination of multiplication and addition. PSI is non-empty if the final value of the decrypted \code{result} is zero.  Note that  the operations on A and B in lines $9$ and $13$-$15$ are homomorphic: $\otimes$ and $\oplus$ correspond to homomorphic multiplication and addition, respectively. Each homomorphic operation adds noise, and thus the noise gets accumulated in the \code{result}. The amount of noise is dependent on the type as well as the number of homomorphic operations: multiplication adds more noise than any other operations. If a ciphertext undergoes a sequence of multiplications, measured in terms of \emph{multiplicative depth}, then it quickly exhausts the noise budget leading to incorrect output due to the noise overflow. Additionally, each homomorphic operation could potentially increase the message size; if this increases beyond the threshold value (plaintext modulus chosen during the initial setup), then the message wraps around leading to a value overflow.

Thus, even though the PSI program is logically correct, the output is not guaranteed to be valid for all inputs. Crucially, it depends on the depth of the circuit (in this case, loop iterations determined by \code{len_A} and \code{len_B}) and the actual inputs.  For example, in SEAL, it fails to execute correctly for larger sets (e.g., \code{len_A} is greater than $4$) due to noise overflows. In OpenFHE and TFHE-rs, it fails when the input values are large due to value overflows. It is thus important to know when the output is valid.

\ila's approach is to statically track the overflows using an expressive type system. Notably, the type system embeds an abstract notion of bounds for the type of polynomials used to in the construction of  ciphertext and plaintexts. It tracks the effect of homomorphic operations on these bounds. The type system constrains the overflows by ensuring that they are always below the threshold values (chosen during the setup). For example, if the product of two ciphertexts with types $\Cipher\ \alpha^1_{\CSort}$ and  $\Cipher\ \alpha^2_{\CSort}$ has the type $\Cipher\ \alpha_{\CSort}$, then the type system computes that $\alpha_{\CSort} = f(\alpha^1_{\CSort}, \alpha^2_{\CSort})$ and checks if $\alpha_{\CSort}$ is less than the threshold values.

Since \ila is parametric in the FHE scheme, each scheme instantiates the abstract bounds with precise information. For example, \ilabgv instantiates the  ciphertext type $\Cipher\ \alpha_\CSort$ with $\cimod{\eps}{\om}{\inf}{\sup}$: intuitively, the ciphertext at level $\om$ has the current noise $\eps$ and the underlying messages lies in the interval $[\inf, \sup]$. Multiplying two ciphertexts with bounds $\code{cipher}$ $\langle  \inf_i, \sup_i,$ $\eps_i, \om \rangle$ yields the new bound $\cimod{\eps}{\om}{\inf}{\sup}$ where $\eps = f(\eps_1, \eps_2)$ and $[\inf, \sup]$ = $[\code{min}(\inf_1, \inf_2), \code{max}(\sup_1, \sup_2)]$. Our formalism leaves $f$ as an abstract function with the only requirement that it is monotonic. This enables one to instantiate \ila in multiple ways by picking the right noise estimator from the existing literature~\cite{norms, ana2, tfhe, ilia2019}. Note that our work itself does not introduce any new noise estimation technique; rather our plug-and-play estimator framework is conducive to user defined estimations. For example, $f(\eps_1, \eps_2) = \eps_1 \times \eps_2$  is a worst-case estimator.  We discuss noise estimators later in Section~\ref{sec:impl}.

\begin{figure}
\begin{tikzcd}[row sep=huge, column sep = huge]
m_1, \dots,m_n \arrow[r, "f" red,  Rightarrow, name=U] \arrow[d, "enc_{pk}", Rightarrow]
& {\color{red}{f}}(m_1, \dots, m_n)  \\
enc(m_1) \dots enc(m_n) \arrow[r, "f" blue, Rightarrow] \arrow[ur, phantom,"\scalebox{1.5}{$\circlearrowleft$}"]
&  {\color{blue}{f}}(enc(m_1), \dots enc(m_n)) \arrow[u, "dec_{sk}", Rightarrow]
\end{tikzcd}
    \caption{\capfig{Functional correctness via commutativity in \ila}{Evaluation on top
    (in {\color{red} red}) is over cleartext messages, while evaluation on
    bottom (in {\color{blue} blue}) is homomorphic over ciphertexts.
    \label{fig:funccorrect}}}
\end{figure}
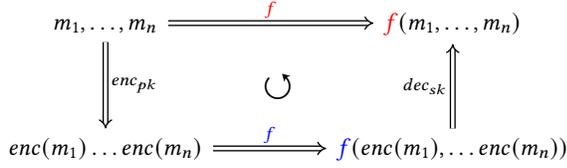

\textbf{Functional correctness.} 
Suppose  the above program passes the type checker, and the circuit executes. Also suppose that  the underlying messages, that is, \code{decrypt_{sk}}(A) and  \code{decrypt_{sk}}(B) are $[m^A_1, m]$ and $[m, m^B_2]$, respectively.  If the type checker is sound, then we should obtain that \code{decrypt_{sk}}(\code{result}) = $0$ (since intersection is non-empty).

Generalizing the above example, we formalize the functional correctness as a
relational property of a pair of executions. The first execution runs the FHE
circuit as intended, whereas the second execution runs the \emph{same} circuit
in \emph{cleartext}, with plaintext inputs that correspond to the ciphertext
ones. We then require that the two executions must match; that is, the first
output decrypts to the second one. In other words, the diagram commutes,
as shown in Figure~\ref{fig:funccorrect}. Theorem~\ref{thm:msg_equiv_commands}
from Section~\ref{sec:ila} states that well-typed \ila programs are
functionally correct (given that the underlying FHE operations are,
with appropriate noise bounds).

\medskip
\noindent \textbf{Noise Management.} 
Going back to the running example, if \ila's type checker rejects the program,
we cannot guarantee that \code{result} will decrypt to $0$. Indeed, the program fails to typecheck in BGV scheme if $len_A$ is $4$. In this case, one has to
use \code{modswitch}, a noise refreshing operation (or PBS in TFHE), to gain additional noise budget. Using the noise budget estimates provided by \ila, line $9$ could be modified to $\code{t4} = \code{result} \otimes \code{t4}$ to gain additional noise budget. If it still fails, depending on the amount of the noise budget either the following operations in lines $13$-$15$ could be switched or line $9$ could be further switched to a lower level in the modulus chain. This process repeats until the type checker accepts or the chain gets fully exhausted. In the latter case, the circuit runs without functional correctness guarantees.

The above iterative process shows that \ila's type checker could be used as a layer for validating noise refreshing transformations. In Section~\ref{sec:msinfer}, we describe a \code{modswitch} inference algorithm, an optimization that tightens the number of homomorphic operations by inferring the effective placement of \code{modswitch} operations in loop-free programs. Specifically,  the inference algorithm  exploits the validation layer to ensure the correctness of the optimization, that is, the optimized program is free of noise overflows. More generally, \ila provides a validation layer for bounds-preserving transformations. In the future, we plan to use the validation layer as a foundation for building correct-by-construction circuit optimizations.

\medskip
\noindent\textbf{Formal \ila Model.}
To summarize, we define \emph{\ila model} using an abstract core language.  Values in the model are categorized into three sorts---message, encoded plaintext, and ciphertext---and each sort is associated with a corresponding set of abstract bounds, such as ciphertext noise and value intervals.  Crucially, the bounds set forms a partial order, and operations over values manipulate these bounds. When the bounds exceed the scheme-specific special elements of the poset, the circuit fails.

%\JG{there was some stuff here about maximal element, but that is not accurate}

We prove that the abstract FHE model enforces functional correctness. Interestingly, instantiating the abstract model with a suitable FHE scheme gives functional correctness for free. This is significant, as any correct instantiation of \ila now enjoys the functional correctness property. Concretely, we instantiate \ila with absolute FHE schemes to obtain  \ilabgv, \ilabfv and \ilatfhe and furthermore prove that they are correct instantiations of \ila.  Thus  programs in instantiated models inherit the functional correctness property. This demonstrates the model's general applicability to RLWE-based FHE schemes.

\section{\ila: A formal model for FHE} \label{sec:ila}
We now present \ila, our core type system for FHE. \ila is parameterized by a
\emph{model}, which encapsulates all necessary semantic details for homomorphic
evaluation. Given an arbitrary model, we then construct a type system for
homomorphic circuits and prove that all circuits constructible under our type
system are functionally equivalent to their non-homomorphic counterparts. 

\subsection{\ila Models}\label{sec:ilamodel}
To use \ila with a particular FHE scheme (e.g., BGV) one needs to define a
\emph{model}, which serves as the semantic interface between the FHE scheme and
\ila's type system. 

\begin{definition}\label{def:ilamodel}
An \ila model $\mathsf{M}$ is given by the following data:
\begin{itemize}
\item Two sets $\ppset$ and $\spset$ for \emph{public} and \emph{secret}
        parameters, respectively, with a mapping $\topub : \spset \to \ppset$;
    \item Sets $\sem{s}$ for each sort $s \in \Sort = \{\MSort, \PSort,
        \CSort\}$ (meaning messages, (encoded) plaintexts, and ciphertexts,
        respectively);
\item Sets  $B_s$ of \emph{bounds} for each sort $s \in \Sort$, partially
    ordered by $\leq_s$;
\item Mappings $|\cdot|^\pparam_\MSort : \sem{\MSort} \to B_\MSort$ and
    $|\cdot|^\pparam_\PSort : \sem{\PSort} \to B_\PSort$, where $\pparam \in \ppset$;
\item A mapping $|\cdot|^\sparam_\CSort : \sem{\CSort} \to B_\CSort$, where
    $\sparam \in
    \spset$;
\item A mapping $\interp^\pparam_\PSort : \sem{\PSort} \to \sem{\MSort}$, where
    $\pparam
    \in \ppset$;
\item A mapping $\interp^\sparam_\CSort : \sem{\CSort} \to
\sem{\MSort},$ where $\sparam \in \spset$;
\item A set of operations $\op$ of arity $\overrightarrow{s_i} \to s$, where
each $s, s_i \in \Sort$, equipped with three mappings:
\begin{itemize}
    \item $\sem{\op} : \overrightarrow{\sem{s_i}} \to \sem{s}$;
    \item A partial mapping $\sem{\op}^\pparam_\bnd : \overrightarrow{B_{s_i}}
        \rightharpoonup B_s$, where $\pparam \in \ppset$;
    \item A mapping on messages $\sem{\op}_\msg : \overrightarrow{\sem{\MSort}} \to
    \sem{\MSort}$, with the appropriate input arity.
\end{itemize}
\end{itemize}
\end{definition}

We now walk through Definition~\ref{def:ilamodel}. 
Each ILA model has a set of public and secret parameters, which 
refer directly to the public and secret keys of
the FHE scheme in question. 
Then, 
each ILA model defines a set of messages, encoded plaintexts, and
ciphertexts, captured through giving semantics $\sem{\cdot}$ to the three sorts $\Sort = \{\MSort, \PSort,
\CSort\}$, respectively. 
%For example, for BGV $\sem{\CSort}$
%
To model quantitative concerns (e.g., ciphertext noise
and integer overflow in the plaintext modulus), each sort $s$ also comes
equipped with a partially ordered set of \emph{bounds}. 
For RLWE-based schemes, 
the bounds for messages and encoded plaintexts will track
integer norms to prevent overflow in the modulus, while the bounds for
ciphertexts will additionally track encryption-induced noise. 
We assume bounds for messages and encoded plaintexts can
be computed using the public key via $|\cdot|_\MSort^\pparam$ and
$|\cdot|_\PSort^\pparam$, while computing bounds for ciphertexts via
$|\cdot|_\CSort^\sparam$ needs the secret
key.

Next, we have \emph{interpretation} functions $\interp_\PSort^\pparam$ and
$\interp_\CSort^\sparam$, mapping their respective sorts to messages. 
The interpretation function for $\PSort$ takes the public parameters as an
argument, while the interpretation function for $\CSort$ takes the secret
parameters. 
Interpreting
encoded plaintexts corresponds to \emph{decoding}, and requires the public parameters, while interpreting
ciphertexts corresponds to \emph{decrypting}, and requires the secret
parameters.
For notational convenience, we define $\interp^\pparam_\MSort$ to
be the identity function, and for $s \in \{\MSort, \PSort\}$, we 
define $\interp^\sparam_s$ to be $\interp^{\topub(\sparam)}_s$.
We omit the sort argument to $|\cdot|$ and $\interp$ when it is clear from
context.

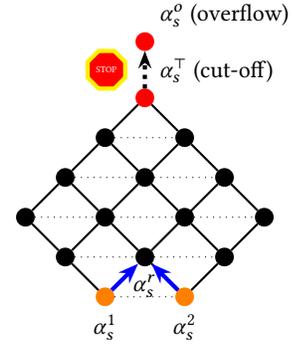
\begin{figure}
\tikzset{dot/.style={fill, circle, inner sep=2.5pt, outer sep=0pt, label=#1}}
\begin{tikzpicture}[node distance=0.5cm]
\node[dot={45:$\alpha^o_s$ (overflow)}, fill=red] (root){};
\node[dot={45:$\alpha^{\top}_s$ (cut-off), below= of root}, fill=red] (top){};
\node[dot={135:}, below left=of top](a){};
\node[dot={45:}, below right=of top](b){};
\node[dot={180:}, below left=of a](c){};
\node[dot={0:}, below right=of a](d){};
\node[dot={0:}, below right=of b](e){};
\node[dot={270:}, below right=of c](f){};
\node[dot={270:}, below right=of d](g){};
\node[dot={270:$\alpha^r_{s}$}, below right=of f](h){};
\node[dot={270:}, below left=of c](i){};
\node[dot={270:}, below right=of e](j){};
\node[dot={270:}, below right=of i](k){};
\node[dot={270:}, below right=of g](l){};
\node[dot={270:$\alpha^1_{s}$}, below right=of k, fill=orange](m){};
\node[dot={270:$\alpha^2_{s}$}, below right=of h, fill=orange](n){};
\draw[thick](top)--(a)--(c)--(f) (a)--(d)--(g) (d)--(f)--(h) (top)--(b)--(d) (b)--(e)--(h);
\draw[thick] (c)--(i)--(k)--(f) (e)--(j)--(l)--(g) (k)--(m) (h)--(n)--(l) (m);
\draw[dotted] (m)--(n) (k)--(h)--(l) (i)--(f)--(g)--(j) (c)--(d)--(e) (a)--(b);
\draw[ultra thick, -{Stealth[scale=0.75]}, blue] (m)--(h);
\draw[ultra thick, -{Stealth[scale=0.75]}, blue] (n)--(h);
%\draw[ultra thick,  -{Stealth[scale=0.75]}, purple] (top)--(root) node[left, midway, regular polygon, regular polygon sides=8, scale=0.5, draw] {stop};
\path[ultra thick, dotted, -{Stealth[scale=0.65]}] (top) edge node[left, midway, regular polygon, regular polygon sides=8, scale=0.5,  draw, yellow, solid,  fill=red, xshift=-0.5cm, scale=0.75]{\color{white}STOP} (root);
\end{tikzpicture}
\caption{A partially-ordered bounds set $B_s$ (indexed by sort $s$).  For any given two points $\alpha^1_s$ and $\alpha^2_s$, the operator bounds mapping $\sem{\op}^\pparam_\bnd$ computes the result bounds $\alpha^r_s$. The operation is well-defined iff $\alpha^r_s \le \alpha^{\top}_s$ where $\alpha^{\top}_s$ is the cut-off for the output to be valid. \label{fig:bounds-poset}}
\end{figure}

Finally, we have a set of operations of varying arity, such as binary homomorphic
addition and multiplication, operations on messages/encoded plaintexts, and
unary \emph{noise management} operations, such as modulus switching.  
Other than its native semantics $\sem{\op}$, each operation comes equipped with 
a partial operation $\sem{\op}^\pparam_\bnd$, operating only on \emph{bounds}, and 
a corresponding \emph{message-level} operation $\sem{\op}_\msg$. 
The operation $\sem{\op}^\pparam_\bnd$ is partial, so that it may throw an error of the
estimated bound (e.g., ciphertext noise) is too large to carry out the
operation. Figure~\ref{fig:bounds-poset} illustrates the operation on bounds.
The message-level operation $\sem{\op}_\msg$ emulates the operator's
intended semantics after  interpretation. For example, the message-level
operation for homomorphic multiplication is multiplication on messages, while 
the message-level operations for unary noise management operations on ciphertexts
are the identity. 

\paragraph{Validity of Models}
A plain ILA model given by Definition~\ref{def:ilamodel} specifies the types of
all operations, but does not encode whether the FHE scheme in question is
\emph{correct}; for example, whether homomorphic addition overflows given
appropriate inputs. To encode correctness conditions for FHE, we require that
the \ila model is additionally \emph{valid}: 
\begin{itemize}[leftmargin=*]
\item \textsf{Commutativity}: for all secret parameters
$\sparam$, 
every $\op : \overrightarrow{s_i} \to s$
and arguments $\overrightarrow{v_i \in \sem{s_i}}$, we have that if
        $\sem{\op}_\bnd^{\topub(\sparam)}(\overrightarrow{|v_i|^\sparam_{s_i}}) = b$, then
$|\sem{\op}(\overrightarrow{v_i})|_s^\sparam \leq_s b$ and
        \[
\interp_s^\sparam(\sem{op}(\overrightarrow{v_i})) =
        \sem{\op}_\msg(\overrightarrow{\interp_{s_i}^\sparam(v_i)}); \] 
\item \textsf{Downwards Closed}:
    If $\op : \overrightarrow{s_i} \to s$, and 
        if $\sem{\op}_\bnd^{\topub(\sparam)}(\overrightarrow{b_i})$ is defined and equals $b$, then
        if $b'_i \leq_{s_i} b_i$ for all $i$, we have that $\sem{\op}_\bnd^{\topub(\sparam)}(
        \overrightarrow{b'_i}) = b'$ with $b' \leq_s b$.
\end{itemize}

The first condition, commutativity, guarantees correctness by requiring that if
the bounds chosen by operator $\op$ are defined and equals $b$, then the output
of $\sem{\op}(\overrightarrow{v_i})$ is equally bounded by $b$, and 
the message-level result of $\sem{\op}(\overrightarrow{v_i})$ given by
$\interp_s^\sparam$ is equal to the value of $\sem{\op}_\msg$ evaluated on the
message-level interpretations of its arguments. 
For homomorphic addition, this amounts to the fact that $\mathsf{enc}(x) \oplus
\mathsf{enc}(y)$ is an encryption of $x + y$; while for bootstrapping, this
condition encodes that $\mathsf{bootstrap}(c)$ does not change the
underlying value inside of $c$. 
The second condition, downwards-closedness, guarantees monotonicity of the
bounds: if an operator states that an
output bound is defined on its inputs, then a smaller output bound is defined on
smaller outputs. 

Throughout the rest of this section, fix an ambient valid \ila model $\mathcal{M}$.
We additionally assume a nullary operator $\mathsf{true} : () \to \MSort$, in
order to evaluate control flow operators.

\subsection{Syntax and Semantics}
Now that we have a model, we now define the core language of \ila. 
The syntax is given in Figure~\ref{fig:ilasyntax}. Expressions may
either be variables, values of sort $\MSort$ or $\PSort$, or applications of
operators. We do not allow hardcoding ciphertexts into programs, as they must
come from the context. Commands are standard for an imperative language.
Note that the core language does not consider loops, while our surface language
(e.g., in Figure~\ref{fig:psi}) does; our concrete type checker only allows
loops which can be statically unrolled with a finite bound. 

\begin{figure}
\small{
\begin{syntax}
\category[Expressions]{e}
\alternative{x}
\alternative{v \in \sem{s}\ (s \in \{\MSort, \PSort\})}
%\\
\alternative{\op(\overrightarrow{e_i})}

\category[Commands]{c}
\alternative{\cskip}
\alternative{x := e}
\alternative{c; c}
\alternative{\eif{e}{c}{c}}
\end{syntax}
}
\caption{\ila Syntax \label{fig:ilasyntax}}
\end{figure}

We give big-step semantics to \ila programs in Figure~\ref{fig:ilastep}. 
Given a substitution $\gamma$ mapping variables to values in $\sem{\MSort} \cup
\sem{\PSort} \cup \sem{\CSort}$, the big-step rules for expressions relate
substitutions and expressions to values. The rule for $\op$ checks that each
input is of the correct sort, and if so, evaluates $\sem{\op}$. 
The big-step rules for commands relate substitutions and commands to output
substitutions, and are standard.

\begin{figure}
\small{
\begin{mathpar}
\judgment{\config{\gamma, e} \Downarrow v} \and 
\inferrule{\gamma(x) = v}{\config{\gamma, x} \Downarrow v}
\and
\inferrule{~}{\config{\gamma, v} \Downarrow v}
\and
\inferrule{\op : \overrightarrow{s_i} \to s \and (\forall i, \config{\gamma,
e_i} \Downarrow v_i
\in \sem{s_i}) \and \sem{\op}(\overrightarrow{v_i}) =
v}{\config{\gamma, \op(\overrightarrow{e_i})} \Downarrow v} 

\\
\judgment{\config{\gamma, c} \Downarrow \gamma'} \and 
\inferrule{~}{\config{\gamma, \cskip} \Downarrow \gamma}
\and
\inferrule{\config{\gamma, e} \Downarrow v}{\config{\gamma, x := e} \Downarrow
\gamma[x \mapsto v]}
\and
\inferrule{\config{\gamma, c_1} \Downarrow \gamma_1 \and \config{\gamma_1, c_2}
\Downarrow \gamma_2}{\config{\gamma, c_1; c_2} \Downarrow \gamma_2}
\and
\inferrule{\config{\gamma, e} \Downarrow \sem{\mathsf{true}}() \and
\config{\gamma, c_1} \Downarrow \gamma'}
{\config{\gamma, \eif{e}{c_1}{c_2}}
\Downarrow \gamma'}
\and
\inferrule{\config{\gamma, e} \Downarrow v \neq \sem{\mathsf{true}}() \and
\config{\gamma, c_2} \Downarrow \gamma'}
{\config{\gamma, \eif{e}{c_1}{c_2}}
\Downarrow \gamma'}
\end{mathpar}
}
\caption{Big-Step Semantics for \ila.\label{fig:ilastep}}
\end{figure}

\subsection{Type System}
\begin{figure}
\centering
\small{
\begin{syntax}
    \categoryFromSet[Bounds]{\alpha_s}{B_s}
    \category[Types]{\tau}
    \alternative{\Cipher\ \alpha_\CSort}
    \alternative{\Plain\ \alpha_\PSort}
    \\
    \alternative{\msg\ \alpha_\MSort}
    \category[Contexts]{\Gamma}
    \alternative{\cdot}
    \alternative{\Gamma, x:\tau}
\end{syntax}
\begin{mathpar}
\centering
    \inferrule{\alpha \leq_s \alpha' \\ s \in \{\mathsf{cipher}, \mathsf{plain},
    \mathsf{msg}\}}{s\ \alpha \leq s\ \alpha'} \and
\end{mathpar}
}
\caption{\ila Types and Subtyping \label{fig:ilatypes}}
\end{figure}

Types for \ila programs, and their associated rules for subtyping, are given in
Figure~\ref{fig:ilatypes}. Each type $\tau$ is a pair
of a sort $s$, and an associated bound $\alpha_s \in B_s$. 
Given a type $\tau$, we write $|\tau|$ for the bound contained in the type.
Subtyping in \ila requires
that the two types are of the same sort and the first associated bound is less
than the second, in the ordering induced by the sort.

\begin{figure}
\small{
\begin{mathpar}
\judgment{\pparam; \Gamma \vdash e : \tau} \and
    \inferrule*[Right=\textsc{var}]{x : \tau \in \Gamma}{\pparam; \Gamma \vdash x : \tau} \and
\inferrule*[Right=\textsc{sub}]{\pparam; \Gamma \vdash e : \tau \and \tau \leq
    \tau'}{\pparam; \Gamma \vdash e : \tau'} \and
    \inferrule*[Right=\textsc{const}]{\sort(v) = s}{\pparam; \Gamma \vdash v : s\ |v|^\pparam_s} \and
    \inferrule*[Right=\textsc{op}]{\op : \overrightarrow{s_i} \to s \and 
(\forall i, \pparam; \Gamma \vdash e_i : \tau_i \wedge \sort(\tau_i) = s_i) \\ 
    \sem{\op}_\bnd^\pparam(\overrightarrow{|\tau_i|}) = \alpha
}{\pparam; \Gamma \vdash 
\op(\overrightarrow{e_i}) : s\ \alpha}
\end{mathpar}
}
\caption{\ila Typing: Expressions \label{fig:ilaexptr}}
\end{figure}

Expressions are typed using Figure~\ref{fig:ilaexptr}. The type checking
judgment takes the 
public parameters $\pparam$ as input to check that all bounds are correct.  
The rule for values computes the bound $|v|_s^\pparam$ as part of the output
type; since we only allow messages and encoded plaintexts in source programs,
we do not need the secret parameters. For operators, we check that all inputs
are sort-correct, and then check that
the output bound 
$\sem{\op}_\bnd^\pparam(\overrightarrow{|\tau_i|})$ is defined; if it is, we record this
in the output type. In essence, since we have a valid \ila model (Section~\ref{sec:ilamodel}),  
the requirement that the output bound is correct enforces that all relevant
quantities (e.g., RLWE noise) are within bounds.

\begin{figure}
\small{
\begin{mathpar}
    \judgment{\pparam; \Gamma \vdash c \dashv \Gamma'} \and
    \inferrule*[Right=\textsc{Assgn}]{\pparam; \Gamma \vdash e : \tau
    }{\pparam; \Gamma \vdash x := e
    \dashv \Gamma[x \mapsto \tau]} \\
    \inferrule*[Right=\textsc{skip}]{~}{\pparam; \Gamma \vdash \cskip \dashv \Gamma} \and 
    \inferrule*[Right=\textsc{seq}]{\pparam; \Gamma \vdash c_1 \dashv \Gamma' \and \pparam; \Gamma' \vdash c_2 \dashv
    \Gamma''}{\pparam; \Gamma \vdash c_1; c_2 \dashv \Gamma''}
\and
    \inferrule*[Right=\textsc{ite}]{\pparam; \Gamma \vdash c_1 \dashv \Gamma_1 \and \pparam; \Gamma \vdash c_2 \dashv
    \Gamma_2 \and \pparam; \Gamma \vdash e : \Msg\
\alpha \and \Gamma_1 \sqcap \Gamma_2 \sim \Gamma'}{\pparam; \Gamma \vdash
    \eif{e}{c_1}{c_2} \dashv \Gamma'}
\and
    \judgment{\Gamma_1 \sqcap \Gamma_2 \sim \Gamma} \and
    \inferrule*{
    (\forall x \in \Gamma, x \in \Gamma_1 \wedge \Gamma_1(x) \leq
    \Gamma(x)) \and 
    (\forall x \in \Gamma, x \in \Gamma_2 \wedge \Gamma_2(x) \leq
    \Gamma(x))
    }{\Gamma_1 \sqcap \Gamma_2 \sim \Gamma}
\end{mathpar}
}
\caption{\ila Typing: Commands \label{appfig:ilacmdtr}}
\end{figure}
Type checking for commands is given in Figure~\ref{appfig:ilacmdtr}, and is largely
standard. 
For \textsf{if} statements, we require that the branch is a cleartext message,
and that both branches type check; then, we \emph{merge} the two output contexts
using the judgment $\Gamma_1 \sqcap \Gamma_2 \sim \Gamma$, which requires that
$\Gamma$ is a sound overapproximation of the post-states represented by
$\Gamma_1$ and $\Gamma_2$.  

\subsection{Semantic Validity}
Before proving full functional correctness, we prove \emph{semantic validity},
which demonstrates that all runtime values are within bounds described by their
types (as modeled by the \ila model). We do so via \emph{semantic type soundness}, which requires we give a
semantics to types: 
\begin{align*}
\sem{\Cipher\ \alpha}^\sparam &= \{ c \in \sem{\CSort} \mid
    |c|^\sparam_\CSort \leq_\Cipher \alpha \} \\ 
    \sem{\Plain\ \alpha}^\sparam &= \{ p \in \sem{\PSort} \mid
    |p|^\sparam_\PSort \leq_\Plain \alpha \} \\
    \sem{\Msg\ \alpha}^\sparam &= \{ m \in \sem{\MSort} \mid |m|^\sparam_\MSort
    \leq_\Msg \alpha \}
\end{align*}
Thus, $\sem{\tau}^\sparam$ are the values of the correct sort that are soundly
approximated by $\tau$ bounds. Note that while we use the secret parameters to
define soundness, only the public parameters $\pparam$ are required for type
checking.

Now, we show that all runtime values of type $\tau$ must reside in
$\sem{\tau}^\sparam$. 
First, given a substitution $\gamma$, we say that 
$\sparam; \Gamma \vDash \gamma$ if for all $x \in \Gamma$, $x \in
\gamma \wedge \gamma(x) \in \sem{\Gamma(x)}^\sparam$.
Then, for expressions, say that $\sparam; \Gamma \vDash e : \tau$ if for all
$\sparam; \Gamma \vDash \gamma$, 
$\config{\gamma, e} \Downarrow v \wedge v \in \sem{\tau}^\sparam$ for some $v$.
Intuitively, $\sparam; \Gamma \vDash \gamma$ states that $\gamma$ is a
well-formed substitution relative to its semantic types, while 
$\sparam; \Gamma \vDash e : \tau$ states that, given a well-formed substitution, 
$e$ evaluates to a well-formed value relative to $\sem{\tau}^\sparam$. 
We now show semantic safety for expressions:
\begin{theorem}[Semantic Safety: Expressions]
Suppose that $\topub(\sparam); \Gamma \vdash e : \tau$. Then $\sparam; \Gamma \vDash e : \tau$.
\end{theorem}
%% \begin{proof}
%%     By induction on the typing derivation. 
%%     For the $\op$ case, we have that each $v_i \in \sem{\tau_i}^\sparam$,
%%     and must show that $\sem{op}(\overrightarrow{v_i}) \in \sem{s\ \alpha}$,
%%     where $\sem{\op}_\bnd^{\topub(\sparam)}(\overrightarrow{|\tau_i|}) = \alpha$.
%%     This follows from commutativity in the model.
%% \end{proof}

For commands, say that $\sparam; \Gamma \vDash c \Dashv \Gamma'$ if for all
$\sparam; \Gamma \vDash \gamma$, 
$\config{\gamma, c} \Downarrow \gamma'$ and $\sparam; \Gamma' \vDash \gamma'$.
Intuitively, commands must map well-formed substitutions under $\Gamma$ to well-formed
substitutions under $\Gamma'$:
\begin{theorem}[Semantic Safety: Commands]
    If $\topub(\sparam); \Gamma \vdash c \dashv \Gamma'$, then we have that 
    $\sparam; \Gamma \vDash c \Dashv
    \Gamma'$.
\end{theorem}
%% \begin{proof}
%%     By induction on the typing derivation. 
%%     The cases for $\mathsf{skip}$, assignment, and sequencing are clear. 
%%     For $\mathsf{if}$, we have that 
%%     $\sparam; \Gamma \vDash c_1 \Dashv \Gamma_1$ and 
%%     $\sparam; \Gamma \vDash c_2 \Dashv \Gamma_2$.
%%     We must show that, if $\Gamma_1 \sqcap \Gamma_2 \sim \Gamma'$ and 
%%     $\sparam; \Gamma_1 \vDash \gamma'$, then 
%%     $\sparam; \Gamma' \vDash \gamma'$, and similarly for $\Gamma_2$.
%%     This follows from observing that if $\sigma \leq \tau$, 
%%     then $\sem{\sigma}^\sparam \subseteq \sem{\tau}^\sparam$.
%% \end{proof}

\subsection{Message Equivalence}
Now that we have proven semantic safety, we prove functional correctness, which
we formalize as \emph{message equivalence}. Intuitively, message equivalence
states that decrypting the result of any homomorphic circuit is equivalent to 
running that same circuit in cleartext, on decrypted inputs. This is the main
guarantee of \ila.

To formalize message equivalence, we give an alternative big-step semantics to
expressions and commands in Figure~\ref{fig:mesteps} that describes cleartext
evaluation. Here, all substitutions
map variables to cleartext values (i.e., values in $\sem{\MSort}$). 
We then define cleartext evaluation by immediately interpreting (i.e., decoding)
all encoded plaintexts into messages, and interpreting operators through their
message-level semantics $\sem{\op}_\msg$. Hence, homomorphic addition will be
interpreted as ordinary addition on integers, and bootstrapping operators will
be interpreted as no-ops. 
The semantics for commands are similar to their ordinary big-step rules.

\begin{figure}
\small{
\begin{mathpar}
\judgment{\config{\gamma, e} \Downarrow^\sparam_\msg v} \and 
\inferrule{\gamma(x) = v}{\config{\gamma, x} \Downarrow^\sparam_\msg v}
\and
\inferrule{v \in \sem{s}}{\config{\gamma, v} \Downarrow^\sparam_\msg
    \interp^\sparam_s(v)}
\and
\inferrule{\op : \overrightarrow{s_i} \to s \and (\forall i, \config{\gamma,
e_i} \Downarrow^\sparam_\msg v_i
) \and \sem{\op}_\msg(\overrightarrow{v_i}) =
v}{\config{\gamma, \op(\overrightarrow{e_i})} \Downarrow^\sparam_\msg v} \and 

\judgment{\config{\gamma, c} \Downarrow^\sparam_\msg \gamma'} \and 
\inferrule{~}{\config{\gamma, \cskip} \Downarrow^\sparam_\msg \gamma}
\and
\inferrule{\config{\gamma, e} \Downarrow^\sparam_\msg v}{\config{\gamma, x := e}
\Downarrow^\sparam_\msg
\gamma[x \mapsto v]}
\and
\inferrule{\config{\gamma, c_1} \Downarrow^\sparam_\msg \gamma_1 \and \config{\gamma_1, c_2}
\Downarrow^\sparam_\msg \gamma_2}{\config{\gamma, c_1; c_2} \Downarrow^\sparam_\msg \gamma_2}
\and
\inferrule{\config{\gamma, e} \Downarrow^\sparam_\msg \sem{\mathsf{true}}() \and
\config{\gamma, c_1} \Downarrow^\sparam_\msg \gamma'}
{\config{\gamma, \eif{e}{c_1}{c_2}}
\Downarrow^\sparam_\msg \gamma'}
\and
\inferrule{\config{\gamma, e} \Downarrow^\sparam_\msg v \neq \sem{\mathsf{true}}() \and
\config{\gamma, c_1} \Downarrow^\sparam_\msg \gamma'}
{\config{\gamma, \eif{e}{c_1}{c_2}}
\Downarrow^\sparam_\msg \gamma'}
\end{mathpar}
}
\caption{Big-Step Semantics for Message Equivalence.}
\label{fig:mesteps}
\end{figure}

We now show that evaluating expressions and commands under their ordinary
semantics, and then interpreting (e.g., decrypting), is equivalent to evaluating
under the alternative semantics $\Downarrow^\sparam_\msg$. 
To define the below theorems formally, we introduce an auxiliary definition: 
given a type context $\Gamma$ and a substitution $\gamma$ such that $\sparam;
\Gamma \vDash \gamma$, let 
\[ \mathsf{interp}^\sparam_\Gamma(\gamma) = \{ x \mapsto
\interp^\sparam_{\sort(\Gamma(x))}(\gamma(x))\}. \]
Thus, $\mathsf{interp}^\sparam_\Gamma(\gamma)$ maps substitutions
$\gamma$ under the original FHE semantics into substitutions only over messages.  
With this definition, we can now prove message equivalence for expressions and
commands:
\begin{theorem}[Message Equivalence: Expressions]\label{def:msgequiv-exp}
    Suppose that $\topub(\sparam); \Gamma \vdash e : \tau$ and $\sparam; \Gamma \vDash \gamma$. 
Then,
    $\config{\gamma, e} \Downarrow v$ and $\config{\interp^\sparam_\Gamma(\gamma), e}
    \Downarrow^\sparam_\msg v_\msg$ such that
    $\interp^\sparam_{\mathsf{sort}(v)}(v) = v_\msg.$
\end{theorem}
The proof  of Theorem~\ref{def:msgequiv-exp} follows by induction on the typing
derivation of $e$. The case for operators makes use of commutativity and
downwards-closedness of the ILA model (Section~\ref{sec:ilamodel}) in order to
interchange decryption and evaluation of the operator. 

After proving message equivalence for expressions, we can then prove it for
commands. The proof is a straightforward induction on the typing derivation for
commands. Appendix~\ref{app:proofs} shows the proof sketch. 
\begin{theorem}[Message Equivalence: Commands] \label{thm:msg_equiv_commands}
    Suppose that $\topub(\sparam); \Gamma \vdash c \dashv \Gamma'$ and $\sparam; \Gamma \vDash \gamma$.
    Then, if 
$\config{\gamma, c} \Downarrow \gamma_1$, we have that 
    $\config{\interp^\sparam_\Gamma(\gamma), c} \Downarrow_\msg \gamma_1'$ such that, for all $x \in
    \gamma_1$, $\interp^\sparam_{\mathsf{sort}(\gamma_1(x))}(
\gamma_1(x)) = \gamma_1'(x)$ .
\end{theorem}

%\section{Connection to BGV scheme} \label{sec:bgvconnection}

\vspace{-0.5em}
\section{\ila Models}\label{sec:ilamodels}
At a high-level, the recipe for scheme instantiation involves defining all sets and mappings from Definition~\ref{def:ilamodel} and proving that the commutativity and downwards axioms hold.

\subsection{\ilabgv: \ila for BGV Scheme} \label{sec:ilabgv}
\begin{figure}
  \[
  \small{
  \begin{array}{l l l l  }
    \hline
     & & \ila & \ilabgv \\
     \hline
     &&&   %n  \in   \mathbb{Z} \quad 
     \eps   \in  \mathbb{Q}_{\ge 1}^+  \\
     &&&\om   \in   \mathbb{Z}^+  \quad
     \inf, \sup  \in  \mathbb{Q} \\
     \hline
     Values  & v ::= & \quad ~\sem{\MSort}  &   \quad~n \in \mathbb{Z}_t^d \\
     &&\mid \sem{\PSort} & \mid {p \in \frac{\mathbb{Z}_t[x]}{(x^d+1)}} \\
%%     &&& \mid \highlight{$\ciphermod{\tht}{\eps}{\inf}{\sup}{\om}$} \\
     \hline
     Expressions & e ::= &  \quad~x \mid v   & \quad~x \mid v  \\
     & & \mid \op(\overrightarrow{e_i}) & \mid e_1 \oplus e_2 \mid e_1 \otimes e_2 \\
     & & & \mid \modswitch{e} \\
     \hline
    Statements & c ::=& \multicolumn{2}{c}{\cskip \mid x := e \mid c; c \mid \eif{e}{c}{c}}\\
    \hline
   Types & \tau ::= & \quad~\Cipher\ \alpha_\CSort  & \quad~ \cimod{\eps}{\om}{\inf}{\sup}  \\
     & & \mid \Plain\ \alpha_\PSort  & \mid \plr{\low}{\high}\\
     & & \mid \msg\ \alpha_\MSort &  \mid \mathbb{Z}_t^d \\
     \hline
  \end{array}
  }
  \]
  \caption{\ilabgv syntax instantiation. Columns \ila and \ilabgv show corresponding syntax. Sorts $\sem{s}$, operator $\op$ and bounds $\alpha_s$ are instantiated accordingly. The row $c$ denoting statements is common to both. \label{fig:ilabgvsyntax}}
\end{figure}

We introduce \ilabgv, an instantiation of \ila model for the BGV scheme, a levelled RLWE scheme~\cite{BGV2012}. Figure~\ref{fig:ilabgvsyntax} shows the syntax correspondence between \ila (Figures~\ref{fig:ilasyntax} and \ref{fig:ilatypes}) and \ilabgv.  Abstract operators, sorts and bounds from \ila are instantiated. Operator $\op$ is instantiated with  homomorphic multiplication $\otimes$,  addition $\oplus$ and modulus switching \code{modswitch}. Plaintext and ciphertext bounds , $\alpha_{\PSort}$ and $\alpha_{\CSort}$, track the interval $[\inf, \sup]$ in which the underlying message lies as well as the current noise $\eps$. Intuitively, $\eps$ measures how far the coefficients are from corrupting the ciphertext. The level $\om$ is drawn from a pre-defined set of levels $\{q_0, q_1, \dots, q_{L} \}$. We say $((\inf, \sup), \eps, \om) \leq_{\CSort} ((\inf', \sup'), \eps', \om)$ if $\inf' \le \inf \le \sup \le \sup'$ and $\eps \le \eps'$.

Since noise $\eps$ is added during encryption, plaintext contains no noise. As a result plaintext-plaintext operations and plaintext ciphertext additions are not noise increasing operations. However, noise increase after a plaintext-ciphertext multiplication is proportional to that of a fresh cipher-ciphertext multiplication. 
%If \code{encode} and \code{decode} are secret parameters, \ilabgv  plaintext values  must include information in addition to polynomial constants. They are thus represented as tuples. Note that the additional information is not required when \code{encode} and \code{decode} are public: one could extract noise and underlying message as a function of the polynomials.
%% Note that ranges are required only for inputs; \ilabgv computes the rest\footnote{Smaller intervals lead to precise overflow detection; however, they may leak information about ciphertexts. The user can turn them off during the actual deployment.}. Cipher bounds also track the current modulus level $\om$ of the ciphertext. This ensures that operands are at the same modulus level.

%% Recall that \ila types carry bounds information $\alpha_{\CSort}$ and $\alpha_{\PSort}$; \ilabgv instantiates these bounds with $(\inf, \sup, \eps, \om)$ and $(\inf, \sup, \eps)$, respectively.
%% The value interval $(\inf, \sup)$ tracks the underlying message interval.

%% \begin{figure}[t]
%%   \[
%%   \input{figs/ilabgv_syntax.tex}
%%   \]
%% \caption{\ilabgv Syntax \label{fig:ilabgvsyntax}}
%% \end{figure}

%All homomorphic operations are parameterized by $\pparam$.

Definition~\ref{def:ilabgv} summarizes the instantiation. Note that it closely follows  \ila model defined in  Section~\ref{sec:ila}.

\begin{definition}[\ilabgv]\label{def:ilabgv}
The \ilabgv model is defined as follows:
\begin{itemize}
%%\item $\spset := \{ \lambda, L, t, d, \chi, \mathsf{S}, B, \ell, \{q_\om\}, pk, sk,relin_k, \code{encode},$ $\code{decode} \}$
\item $\spset := \{  t, d, \{q_\om\}, pk, relin_k, \code{encode},\code{decode} \} \times \{ sk \}$
\item $\ppset = \{ d,  t,\{q_\om\}, pk, relin_k, \code{encode}, \code{decode} \}$
\item $\topub = \spset = \ppset \times \{ sk \} \to \ppset$ is the projection onto $\ppset$
\item Sort $\Sort = \{\MSort, \PSort, \CSort\}$
  \begin{itemize}
  \item $\sem{\MSort} = n \in \mathbb{Z}_t^d$
  \item $\sem{\PSort} = p \in \frac{\mathbb{Z}_t[x]}{(x^d+1)} $
    \item $\sem{\CSort} = ct \in \cring{q}{n+1} \times \mathbb{Z}_{\ge 0} $
  \end{itemize}
\item Encoding and decoding functions:
  \begin{itemize}
  \item \code{encode}: $\mathbb{Z}^d_t \to \ring{t}$
  \item \code{decode}: $\ring{t} \to \mathbb{Z}^d_t$
    \end{itemize}
\item Bounds sets:
 \begin{itemize}
 \item $B_{\MSort} = \emptyset$
  \item $B_{\PSort} = \{(\inf, \sup) \mid \inf \le \sup  \}$ 
    ordered by $\leq_{\PSort}$
    \item $B_{\CSort} = \{(\inf, \sup, \eps, \om) \mid \inf \le \sup \text{ and } \eps, \om \in \mathbb{N} \}$ 
    ordered by $\leq_{\CSort}$
    \end{itemize}
\item Mappings for bounds computation:
\begin{itemize}
  \item $|m|^\pparam_\MSort = \emptyset $
%  \item $|\plainr{p}{\low}{\high}{\eps}|^\pparam_\PSort = (\inf, \sup, \eps) $
  \item $|p|^\pparam_\PSort = (\inf, \sup)$ (explained below)
\item  $|\tht, \om|^\sparam_\CSort  = (\inf, \sup, \eps, \om) $ (explained below)
\end{itemize}
\item Decoding on plaintexts: $\interp^\pparam_\PSort(p) = \code{decode}(p)$
\item Decryption on ciphertexts: $\interp^\sparam_\CSort(\tht) = \code{decode}(\code{decrypt}(\tht))$
\item Operator $\op \in \{ \oplus, \otimes, \code{modswitch} \}$ defined as follows:
\begin{itemize}
 \item $\sem{\otimes}$: $\{\CSort, \PSort\}$, $\{\CSort, \PSort \}$ $\to$ $\{\CSort, \PSort \}$
 \item $\sem{\oplus}$: $\{\CSort, \PSort\}$, $\{\CSort, \PSort \}$ $\to$ $\{\CSort, \PSort \}$
% \item  \textbf{Scalar multiplication.} $\sem{\times}$: $\integer, \CSort$ $\to$ $\CSort$
 \item $\sem{\code{modswitch}}$: $\CSort$ $\to$ $\CSort$
 \end{itemize}
 \item Operator bounds manipulation mapping, $\sem{\op}^\pparam_\bnd$ defined in Table~\ref{tab:operations-bgv}.
\end{itemize}
\end{definition}

The set $\spset$ contains all scheme parameters including plaintext modulus $t$, poly modulus degree $d$, scheme-related keys as well as encoding and decoding functions. The public parameter set $\ppset \subseteq \spset$ contains all public parameters including public and relinearization keys. The function $\topub$ projects $\spset$ to $\ppset$.

Bounds sets for plain and ciphertexts include set of elements, $\alpha_{\PSort}$
and $\alpha_{\CSort}$ (described above). Bounds set for messages is empty. In
bounds computations for plaintext $p$ if $z = \code{decode}(p)$ then $\inf$ and $\sup$ are two random rationals such that $\inf \le \min_i \{ z_i \} \le \max_i \{ z_i \} \le \sup$. Bound computations for ciphertext involves decrypting the ciphertext using the secret key to obtain minimum and maximum of the underlying cleartext. If $\code{eval\_noise}_{sk}$ computes noise in a ciphertext then $\eps$ is chosen (deterministically or randomly) such that  $\code{eval\_noise}_{sk}(\tht) \le \eps$. We also assume (but not explicitly define) the decoding and decryption functions. Though these are BGV scheme primitives, they are outside \ilabgv model as these operations are typically used at the end of the circuit evaluation.

\begin{table*}
  \scalebox{0.9}{
    \begin{small}
\begin{tabular}{@{}lllll@{}}
    \toprule
    \multicolumn{1}{@{}l}{\textbf{Op}} &
    \multicolumn{2}{c}{$\sem{\cdot}^\pparam_\bnd $} &
    \textbf{Conditions}\\[2pt]
    \bottomrule
    %%%
     & $\sem{\oplus}^\pparam_\bnd $($\overrightarrow{\inf_i, \sup_i, \eps_i, \om}$) & =~ $(\inf, \sup, \eps, \om)$ & if $-t/2 \le \inf \le \sup < t/2$ and $\eps \le \kappa$\\
    $\oplus $ &$\sem{\oplus}^\pparam_\bnd $($\overrightarrow{\inf_i, \sup_i}$)   &=~ $(\inf, \sup )$ & if $-t/2 \le \inf \le \sup < t/2$\\
     &$\sem{\oplus}^\pparam_\bnd $($(\inf_1, \sup_1)$, $(\inf_2, \sup_2, \eps_2, \om)$) &=~ $(\inf, \sup, \eps, \om)$ & if $-t/2 \le \inf \le \sup < t/2$ and $\eps \le \kappa $\\
    &&& \qquad where \\
    &&& $\inf := \inf_1 + \inf_2$, $\sup := \sup_1 + \sup_2$, $\eps := \eps_1 + \eps_2$ \\
    \hline
    & $\sem{\otimes}^\pparam_\bnd $($\overrightarrow{\inf_i, \sup_i, \eps_i, \om}$) &=~ $(\inf, \sup, f(\eps_1, \eps_2), \om)$ & if $-t/2 \le \inf \le \sup < t/2$ and $f(\eps_1, \eps_2) \le \kappa$\\
    $\otimes$ &  $\sem{\otimes}^\pparam_\bnd $($\overrightarrow{\inf_i, \sup_i}$)  &=~ $(\inf, \sup)$ & if $-t/2 \le \inf \le \sup < t/2$ \\
     &$\sem{\otimes}^\pparam_\bnd $($(\inf_1, \sup_1)$, $(\inf_2, \sup_2, \eps, \om)$)&=~ $(\inf, \sup, g(\eps), \om)$ & if $-t/2 \le \inf \le \sup < t/2$ and $g(\eps) \le \kappa$\\\
    &&& \qquad where \\
        &&& {$\inf := \min \{{\inf}_1 * {\inf}_2, {\inf}_1 * {\sup}_2 , {\sup}_1 * {\inf}_2, {\sup}_1 * {\sup}_2 \}$ }\\
        &&& {$\sup := \max \{{\inf}_1 * {\inf}_2, {\inf}_1 * {\sup}_2 , {\sup}_1 * {\inf}_2, {\sup}_1 * {\sup}_2 \}$ }\\
\hline
%% $\times$  & $\sem{\times}^\pparam_\bnd $($n,(\inf, \sup, \eps, \om)$) &=~ $(\inf', \sup', \eps', \om)$ & if $-t/2 \le \inf' \le \sup' < t/2$ and $\eps' \le l_\om$\\
%% &&& where  $\inf' := n* \inf$, $\sup' := n * \sup$, $\eps' := n * \eps$ \\
%% \hline
$\code{modswitch}$  &$\sem{\code{modswitch}}^\pparam_\bnd(\inf, \sup, \eps, \om) $ &=~ $(\inf, \sup, \eps', \om-1)$ & $0 \le \om-1 \le L-1 $; $\eps' = \frac{q_{\om-1}}{q_\om}\eps + B_r \le l_{\om-1}$ \\
\bottomrule
\end{tabular}
\label{tab:operators}
\end{small}

 }
\caption{\ilabgv: \ila model for BGV and BFV scheme illustrating key homomorphic operations.
  Constant $\kappa$ is $\frac{q_{\om}}{2}$ for BGV, and $\frac{1}{2}$ for BFV; scheme parameters $t$ and $q_\om$ are  plaintext modulus and coefficient modulus at modulus level $\om$, respectively.}

\label{tab:operations-bgv}
\end{table*}

 Table \ref{tab:operations-bgv} provides the bounds computations for the key \ilabgv homomorphic operations. The second column computes the operator bound $\sem{.}^{\pparam}_{\bnd}$ and the last column describes the conditions under which the bound is defined.
 For example, $\tht_1 \oplus \tht_2$ with input bounds $(\inf_i, \sup_i, \eps_i, \om)$ results in the bound $(\inf_1+\inf_2, \sup_1+\sup_2, \eps_1+\eps_2, \om)$ provided $-t/2 \le \inf \le \sup < t/2$ and $\eps \le \frac{q_\om}{2} $ where $t$ and $q_{\om}$ are the plaintext modulus and critical value at modulus level $\om$, respectively. Operator \code{modswitch} decrements the modulus level by one and reduces the original noise by $\frac{q_{\om-1}}{q_{\om}}$; a small correction $B_r$ is also added to correct the rounding error induced by the division.
 
 Our noise estimation for homomorphic multiplication is parametric in functions $f$ and $g$ with the only restriction that they are downwards closed. Thus,  they are positive and monotonic. In Section~\ref{sec:impl}, we refer to estimators from the existing literature that satisfy these conditions.

\subsubsection{\ilabgv Typing Rules}
\ilabgv typing rules are obtained by instantiating \ila type system from Figure~\ref{fig:ilaexptr} using the bounds information computed in Table~\ref{tab:operations-bgv}. 
 Below, we show a sample typing rule for $\otimes$ operator that multiplies two ciphertexts.
%% On the left, we show the \ila operator typing rule and on the right the corresponding instantiation in \ilabgv.

\vspace{-1em}
\begin{mathpar}
  \inferrule{
  \otimes : \CSort, \CSort \to \CSort \\
  \forall i, \jdg {\Ga} {e_i} {\cimod  {\eps_i}{\om}{{\inf}_i}{{\sup}_i}} \\
    -t/2 \le {\inf} \le {\sup}  < t/2 \\
    {f(\eps_1, \eps_2)\le q_{\om}/2}}
    {\jdg {\Ga} {e_1 \otimes e_2} {\cimod {f(\eps_1 , \eps_2)}{\om}{\inf}{\sup}}}
    \end{mathpar}

The output interval $[\inf, \sup]$ is computed from Table~\ref{tab:operations-bgv}. The exact definition of $f$ is dependent on the noise estimation technique. This enables an extensible framework for plug-and-play noise estimators. Appendix~\ref{app:ilabgvtr} shows typing rule instantiation for \code{modswitch}.

%\input{ila-bgv}

%\subsection{\ilabgv Type System}
%\input{ila-bgv-types}

%% \medskip
%% We complete the instantiation by proving that commutativity and downwards closure axioms hold for \ilabgv.

%\subsubsection{Instantiation of \ila axioms}

\subsubsection{Instantiation of \ila Functional Correctness} 

To complete the \ilabgv instantiation, we have to show that commutativity and downwards closure axioms hold in our instantiation. Downwards closed axiom is immediate as the noise growth in homomorphic operations is monotonic.

The commutativity axiom partly encodes the correctness theorems of the BGV scheme. Intuitively, it says that the cipher addition of two $\om$ level ciphers with noise $\eps_1$ and $\eps_2$ is a valid ciphertext i.e., decryption yields a \emph{correct} plaintext value when $\eps_1 + \eps_2 \le q_\om/2$. Similarly, the cipher multiplication of such ciphertexts is valid when $\eps_1 * \eps_2 \le q_\om/2$ (cf. Lemma 7 and 8 from Section 5~\cite{BGV2012}). Further, $-t/2 \le \inf_1 + \inf_2 \le \sup_1 + \sup_2 < t/2$ guarantees that the resultant plaintext decodes correctly. Correctness up to plaintext space follows Lemma 7 and 8 from Section 5~\cite{BGV2012}. The rest is immediate by induction of typing derivation.

Since \ilabgv is a valid \ila model, we get Theorem~\ref{thm:msg_equiv_commands} for free. That is if the initial environment contains correct-by-construction ciphertexts, then
well-typed \ilabgv programs are functionally correct (up to the correctness of FHE primitive operations). Specifically, the programs are devoid of noise overflow errors. Additionally, the decryption and decode of the output(s) exactly matches the corresponding cleartext values, i.e., the absence of any plaintext modulus wraparound errors. We thus guarantee the functional correctness of a BGV circuit via type checking.

Note the assumption on the initial environment. If the ciphertexts are adversarial, i.e., they do not match the input types, then the theorem guarantees are invalid. Thus, the input sources are trusted. This is a reasonable assumption as FHE circuits run on sensitive but trusted inputs.

\subsection{\texorpdfstring{\ilabfv: \ila}{} for BFV Scheme} \label{sec:ilabfv}
BFV is a levelled RLWE scheme that is highly similar to BGV except that messages are scaled up and encoded from the most significant bit.
\ilabfv is the instantiation of \ila for BFV scheme.
For brevity, we only focus on key differences.

In a BFV scheme modulus switching is used as a tool to improve computational performance.
Since this is orthogonal to the current focus, \ilabfv supports all operations of \ilabgv except  \code{modswitch}. 
Thus, the instantiation is similar to that of \ilabgv  with two differences. 
First, the  cipher bounds set $B_{\CSort}$ is the set of triples, $(\inf, \sup, \eps)$ with the standard lexicographic ordering. 
Note that modulus level is not included. Second, the operator bounds manipulation mapping uses the condition $\eps \le \frac{1}{2}$ instead of $\eps \le \frac{q_{\om}}{2}$ (the last column of Table~\ref{tab:operations-bgv}).

It is obvious that \ilabfv also satisfies commutativity and downwards closed axioms, and thus we  get correctness, that is, Theorem~\ref{thm:msg_equiv_commands}, for free.
%\subsubsection{Instantiation of \ila axioms}
%\subsubsection{Instantiation of \ila Functional Correctness} 

\subsection{\texorpdfstring{\ilatfhe: \ila}{} for TFHE Scheme} \label{sec:ilatfhe}
%\subsubsection{Instantiation of \ila axioms}
We introduce \ilatfhe, an instantiation of \ila model for the CGGI/TFHE scheme~\cite{tfhe}.
 TFHE is a fast bootstrapping based scheme. It differs from BGV in that it supports additional types of ciphertexts (\lwe, \rlwe and \gsw) and the corresponding operations including boolean (e.g., exclusive-or). Plaintext and ciphertexts are elements of the sets $\ring{t}$ and $\mb_{q}^{n+1} \cup \ring{q}^{n+1}$, respectively (see Section~\ref{sec:background}). \lwe and \rlwe ciphertexts are drawn from the sets $\mb_{q}^{n+1}$ and $\ring{q}^{n+1}$, respectively; a \glwe ciphertext is either a \lwe cipher or a \rlwe cipher. A \gsw ciphertexts can be seen as elements of the space $\ring{q}^{{n+1}\times {n+1}}$. 
 
 \gsw ciphertexts should be considered as tools necessary to support various internal TFHE operations that are not exposed to the user. Hence, a ciphertext is an element of either $\mb_{q}^{n+1}$ or $\ring{q}^{n+1}$, and we track the set using $id$. Thus, we define $\alpha_{\CSort}$ as $(id, \inf, \sup, \eps)$.
We say $(id, \inf, \sup, \eps) \leq_{\CSort} (id', \inf', \sup', \eps')$ if $id = id'$ followed by the standard lexicographic ordering of the remaining triples.
The bounds set for plaintext is the same as that of BGV. Similarly, a plaintext could be an element of $\mb_{t}$ or $\ring{t}$, however the former coincides with the cleartext space. As a result we assume a plaintext is always a RLWE plaintext.

TFHE supports programmable bootstrapping (PBS), \code{pbs}, for noise reduction in the ciphertext. Recall from Section~\ref{sec:background} that programmable bootstrapping is a composition of modulus switching, blind rotation and sample extraction operations. For PBS to succeed, the switching operation must succeed; the latter requires a non-zero noise budget.

TFHE also supports internal product, $\boxtimes$ that  operates exclusively on \glwe ciphertexts and an external product, $\boxdot$, that operates on (restricted) combinations of ciphertexts. TFHE supports more operators, however, these additional operators are implemented as a combination of basic operators. For example, \underline{c}ontrolled \underline{mu}ltiple\underline{x}er, \code{cmux}, that conditionally selects a ciphertext is implemented as a combination of external product, \glwe cipher addition and multiplication. Definition~\ref{def:ilatfhe} summarizes the key  details and focuses on differences with respect to the BGV scheme. We provide the full instantiation in Appendix~\ref{app:tfhe}.

%\vspace{-1em}
\begin{definition}[\ilatfhe]\label{def:ilatfhe}
  The \ilatfhe model is defined as follows:
\vspace{-0.5em}
  \begin{itemize}
  \item Sort $\Sort = \{\MSort, \PSort, \CSort\}$
  \begin{itemize}
  \item $\sem{\MSort} = n \in \mathbb{Z}_t$
  \item $\sem{\PSort} = p \in \frac{\mathbb{Z}_t[x]}{(x^d+1)} $
    \item $\sem{\CSort} = ct \in \cring{q}{n+1} \cup  \mathbb{Z}^{n+1}_t $
  \end{itemize}
  \item Bounds sets:
   \begin{itemize}
      \item $B_{\CSort} = \{(\id, \inf, \sup, \eps) \mid \inf \le \sup, \\ \id \in \{ \lwe, \rlwe, \gsw \} \text{ and } \eps\in \mathbb{N} \}$ 
      \end{itemize}
  \item Mapping for bounds computation:
  \begin{itemize}
    \item  $|\tht|^\sparam_\CSort  = (\id, \inf, \sup, \eps) $ (explained below)
  \end{itemize}
  \item Operator $\op \in \{ \oplus, \otimes,  \boxtimes,  \boxdot, \code{pbs} \}$. Select operators defined as follows:
  \begin{itemize}
   \item $\sem{\boxdot}: \{\CSort, \CSort\} \to \{\CSort \}$
   \item $\sem{\code{pbs}}:$ $\{\CSort, \CSort \}$ $\to$ $\{\CSort \}$
   \end{itemize}
   \item Operator bounds manipulation mapping, $\sem{\op}^\pparam_\bnd$ defined in Table~\ref{tab:tfhe_operations}.
  \end{itemize}
  \end{definition}

Similar to \ilabgv, \ilatfhe defines scheme parameters as well as bounds sets. For a cipher $ct$ the map $|.|^\sparam_\CSort$ is defined as a tuple $ (\id, \inf, \sup, \eps) $, where $\id = \lwe$ if the coefficients of the given cipher are elements of $\mathbb{Z}_t$ and $\id = \rlwe$ if the coefficients are elements of $\ring{q}^{n+1}$. Value bounds $\inf$ and $\sup$ are random rational numbers such that $\inf \le \code{decrypt}(ct) \le \sup$. The noise in $\tht$ is bounded by $\eps$ i.e., $\code{eval\_noise}_{sk}(ct) \le \eps$.
Table~\ref{tab:tfhe_operations} defines the operator bounds manipulation mappings for select TFHE operations. Despite the differences in the types of ciphertexts, the bounds manipulation is mostly straightforward for all types of ciphertext multiplication. The output noise is parameterized by the functions that are downward closed; the operation is defined iff the output noise less than $\frac{q}{t}$. 

Given two ciphertexts $ct_1$ and $ct_2$, the $\code{pbs}$ operator applies the function (e.g., LUT) in $ct_1$ to the second input $ct_2$. At the same time it resets the noise in the output to the nominal level (constant noise denoted by $\eps_b$). The operation emits a \lwe ciphertext with new bounds provided there are no overflows. The condition $-t/2 < \inf_0 < \sup_0$ ensures that there is no overflow in the output value. Note that it is sufficient to check that $\inf_0$ and $\sup_0$ as they are the value bounds of the LUT in $ct_1$. We can ignore $\inf$ and $\sup$ from the input $ct_2$. Interestingly, the constraint on the input noise $\eps < q/t - \delta$ is required to ensure that the switching step of the bootstrapping operation succeeds (see Section~\ref{sec:background}); $\delta$ is a constant representing switching error, thus $\eps$ must have a non-zero noise budget for PBS to succeed.

\begin{table*}
\scalebox{0.9}{

\begin{small}
    \begin{tabular}{@{}llll@{}}
        \toprule
        \multicolumn{1}{@{}c}{\textbf{Op}} &
    %    \multicolumn{2}{@{}c}{\textbf{Signature }} &
        \multicolumn{1}{c}{$\sem{\cdot}^\pparam_\bnd $} &
        \textbf{Conditions}\\[2pt]
        \bottomrule
        %%%
        \hline
        $\sem{\otimes}^\pparam_\bnd $($\overrightarrow{\lwe, \inf_i, \sup_i, \eps_i}$) &=~ $(\lwe, \inf, \eps_b)$ & $\eps_b $ is nominal ($=1$)\\
        \hline
        $\sem{\boxtimes}^\pparam_\bnd (\overrightarrow{\gsw, \inf_i, \sup_i, \eps_i})$ &=~ $(\gsw, \inf, \sup, f(\eps_1, \eps_2))$ & $f(\eps_1, \eps_2) \le q/t$\\
        \\
        $\sem{\boxdot}^\pparam_\bnd (\gsw, \inf_1, \sup_1, \eps_1), (\id, \inf_2, \sup_2, \eps_2)$ &=~ $(\rlwe, \inf, \sup, g(\eps_1, \eps_2))$ & $g(\eps_1, \eps_2) \le q/t$\\
        \hline
        \\
    $\sem{\code{pbs}}^\pparam_\bnd \big ( (\gsw, \inf_0, \sup_0, \eps_0), (\lwe, \inf, \sup, \eps) \big )$ &=~ $(\lwe, \inf, \sup, \eps_b )$ & $ \max\{ \log_2|\inf|, \log_2 |\sup|\} \le (\log_2(t) -1)$\\
    && $\eps < q/t - \delta$ and $-t/2 \le  \inf_0 \le  \sup_0 < t/2$ \\
        %%%
    \bottomrule
    \end{tabular}
    \label{tab:tfhe_ops}
    \end{small}

}
\caption{\ilatfhe: \ila model for TFHE scheme illustrating select TFHE operations.
 Input ciphertext operand has the type $(id_i, \inf_i, \sup_i, \eps_i)$ }
\label{tab:tfhe_operations}
\end{table*}

The commutativity and downwards closure axioms are similar to that of \ilabgv.
Since \ilatfhe is a valid \ila model, we get Theorem~\ref{thm:msg_equiv_commands} for free.

\vspace{-0.5em}
\section{Transformation Validation}\label{sec:msinfer}

%\ila not only enables  correct-by-construction circuits but also correct-by-construction circuit transformations.
\ila's type system can be used to validate that the transformations are bounds-preserving. For example,   transforming $x_1 \otimes x_2 \otimes x_3 \otimes x_4$ to $y_1 \otimes y_2$ where $y_1 = x_1 \otimes x_2$ and $y_2 = x_3 \otimes x_4$ is not bounds-preserving: the former has a multiplicative depth of $3$ where as the latter has the multiplicative depth of $2$; this leads to different noise levels in the output. This is useful in building correct-by-construction optimizations.

We describe an optimization that infers the  placement of modulus switching operations for loop-free programs and uses the type system to validate the bounds-preservation property.

\begin{algorithm}
\caption{Modulus Switch (MS) Inference \label{alg:msinfer}}
\SetKwFunction{FcreateDefList}{DefList}
\SetKwFunction{FMSinsert}{MSinsert}
\SetKwFunction{FMSLevel}{MSLevel}
\SetKwFunction{FMulDepthTree}{MulDepthTree}
\SetKwFunction{FTypeCheck}{TypeCheck}
\SetKwFunction{FModswitch}{Modswitch}
\SetKwProg{Fn}{Function}{:}{}

 \tcp{Create a definition list.}
 $\texttt{defList} \gets \FcreateDefList(\texttt{ast})$\;
\tcp{Create a MD Tree for a given \texttt{node}.}
$\texttt{mdtree} \gets \FMulDepthTree{\texttt{node}, $\phi$}$\;
$\texttt{ast} \gets \FModswitch{\texttt{ast}, \texttt{mdtree}, $\Gamma$}$\;

\Fn{\FModswitch(\texttt{ast}, \texttt{mdtree}, $\Gamma$) $\rightarrow$  \texttt{AST} }{
       \tcp{Pick a leaf node.}
       $\texttt{node} \gets \texttt{Leaves(mdtree)}$\;
       $\texttt{p} \gets \texttt{Parent(node, mdtree)}$\;
       $\texttt{rhs} \gets \texttt{defList}[\texttt{p}]$\;
       $\texttt{ast}, \texttt{rhs'} \gets \FMSinsert(\texttt{ast}, \texttt{p}, \texttt{rhs})$\;
       $\_, \tau \gets \FTypeCheck{\texttt{rhs'}, $\Gamma$}$\;
       $\Gamma \gets \Gamma[\texttt{node} \mapsto \tau]$\;
%%       $\Gamma \gets \Gamma'$\;
       $\texttt{node} \gets \texttt{p}$\;
       \While{\texttt{node} $\ne$ \texttt{root(ast)}}{
       $\texttt{p} \gets \texttt{Parent(node, mdtree)}$\;
       $\texttt{rhs} \gets \texttt{defList}[\texttt{p}]$\;
       $\texttt{ast}, \texttt{rhs} \gets \FMSLevel(\texttt{ast}, \texttt{p}, \texttt{rhs}, \Gamma)$\;
       $\texttt{stat}, \tau \gets \FTypeCheck{\texttt{rhs}, $\Gamma$}$\;
       \If{!\texttt{stat}}{
         \Return fail\;
         }
       $\Gamma \gets \Gamma[\texttt{node} \mapsto \tau]$\;
  %%     $\Gamma \gets \Gamma'$\;
       $\texttt{node} \gets \texttt{p}$\;
       }
       %% $\texttt{status}, \_ \gets \FTypeCheck{\texttt{rhs}, $\Gamma$}$\;
       %% \If{$!\texttt{status}$}{
       %%   $\texttt{rhs'} \gets \FMSinsert(\texttt{p}, \texttt{rhs})$\;
       %%   $\_, \tau \gets \FTypeCheck{rhs', $\Gamma$}$\;
       %%   $\Gamma' \gets \Gamma[\texttt{node} \mapsto \tau]$\;
       %%   $\Gamma \gets \Gamma'$\;
       %%   }
       \Return ast\;
}

\tcp{Type check an expression.}
\Fn{\FTypeCheck(e, $\Gamma$) $\rightarrow$ (\texttt{bool}, $\tau$)}{
$\texttt{status}, \tau \gets \dots$\;
\Return \texttt{status}, $\tau$\;
}

\tcp{Ensure that  operands have same modulus level.}
\Fn{\FMSLevel(\texttt{ast}, \texttt{p}, \texttt{rhs}, $\Gamma$) $\rightarrow$ (\texttt{bool}, $\tau$)}{
$\texttt{ast}, \texttt{rhs} \gets \dots$\;
\Return \texttt{ast}, \texttt{rhs}\;
}

%% \Fn{\FcreateDefList{$AST$}}{
%% $status \gets true$\;
%% }

\Fn{\FMulDepthTree{\texttt{node}, \texttt{wl}}}{
       $\texttt{rhs} \gets \texttt{defList}[\texttt{node}]$\;
       \tcp{Collect  multiplicative operands}
       $\texttt{wl} \gets \texttt{wl} \cup \texttt{Mulops(rhs)} $\;
       \ForEach{$\texttt{xnode} \in \texttt{wl}$}{
            \If{$\texttt{xnode} \in \texttt{dom(defList)}$}{
                  $\texttt{xtree} \gets \FMulDepthTree{\texttt{xnode}, $\phi$}$\;
                  \tcp{Update children set for \texttt{node}}
                 $\texttt{node.child} \gets \texttt{node.child} \cup \texttt{xtree}$\;
            }
       }

    \tcp{returns MD Tree rooted at \texttt{node}}
       \Return \texttt{node}\; 
}

\end{algorithm}

\subsection{Modulus Switch Inference}
Recall from Section~\ref{sec:background} that modulus switching refreshes noise. Though it reduces  both noise and noise budget in a ciphertext, the relative reduction in noise is greater than that of noise budget. This is significant and somewhat surprising: if a program turns on switching level for every ciphertext, then the noise budget gets quickly exhausted reducing the circuit's overall multiplicative depth (compared to the vanilla circuit). Moreover, the operation is computationally expensive.

However, modulus switching could improve the circuit's multiplicative depth. Our inference algorithm identifies the conditions under which modulus switch placement succeeds. Toward this end, the algorithm iteratively estimates the placement and uses the type checker to validate the placement at every iteration.

\begin{figure}[t]
\begin{minipage}{0.18\textwidth}
\begin{lstlisting}[language=Python, style=mystyle, mathescape=true]
c2 = c1 $\otimes$ c1
c3 = c2 $\otimes$ c2
c4 = c3 $\otimes$ c3
c5 = c4 $\otimes$ c4
\end{lstlisting}
\end{minipage}
\begin{minipage}{0.29\textwidth}
\begin{lstlisting}[language=Python, style=mystyle, mathescape=true]
c2 = c1 $\otimes$ c1
c3 = modswitch(c2) $\otimes$ modswitch(c2)
c4 = c3 $\otimes$ c3
c5 = c4 $\otimes$ c4
\end{lstlisting}
\end{minipage}
\caption{Modulus switch inference on the left program yields the right program. \label{fig:msinfer}}
\end{figure}

Consider the program shown on the left of Figure~\ref{fig:msinfer}. It computes $\code{c1}^{16}$. Suppose that $\code{c5}$ in line $4$  yields incorrect output due to the noise overflow. Transforming line $4$ to  $\code{c5} = \code{modswitch}(\code{c4} \otimes \code{c4})$ will not help as switching cannot recover any additional noise budget after-the-fact.
So where should the  placement be such that the computation succeeds? Our novel insight is to place the switching  operations at \emph{least positive multiplicative depth}. This yields maximal noise budget savings that could enable more operations. In this case, the multiplicative depth of \code{c5} is $15$ (since \code{c5} = $\code{c1}^{16}$ with $15$ multiplications) and that of \code{c2} is $1$. Thus,  switching \code{c2} yields maximal noise budget gains.

However, there are two alternatives. First, changing line $1$ to  $\code{c2} = \code{modswitch}(\code{c1} \otimes \code{c1})$; this causes the modulus level of \code{c2} to be one less than that of \code{c1}. Second, changing line $2$ to $\code{c3} = \code{modswitch(c2)} \otimes \code{modswitch(c2)}$; this leaves the level of \code{c2} unchanged.

Recall that switched operands can only operate with operands at the same level (Section~\ref{sec:background}). Thus, the former approach initiates a modulus level propagation chain: every use of \code{c2} must ensure that the remaining operands are also at the same modulus level; otherwise the operation fails. However, the latter approach leaves the level of \code{c2} unchanged;
%it is equivalent to replacing every occurrence of \code{modswitch(c2)} with \code{c2'}. Our algorithm chooses the latter approach.
this allows program to use  \code{c2} as much as possible minimizing the number of changes.

Algorithm~\ref{alg:msinfer} presents the modulus switch (MS) inference algorithm. The input abstract syntax tree (AST) is assumed to be in Static Single Assignment (SSA) form, a reasonable assumption, as many widely used compiler intermediate representations adopt SSA to facilitate optimization.

At a high level, the algorithm has three key steps. First, using the function \texttt{DefList}, it  constructs a list of  variable definitions from the program. Second, using the function \texttt{MulDepthTree}, it constructs a  multiplicative depth tree (MD tree) for the variable (representing ciphertext) that fails to type check due to  noise overflow. Third, using function \texttt{Modswitch}, it  inserts switching operations along the path with the highest multiplicative depth.

The function \texttt{MulDepthTree} constructs the MD tree. The multiplicative depth is defined as the maximum distance from the root to any leaf node. Given a node \texttt{node}, the tree is built by recursively identifying the multiplicative operands that contribute to \texttt{node} and representing them as its child nodes. Given an expression $e$, the function \texttt{Mulops} returns the set of variables that appear as multiplicative operands---either directly in $e$, or within any of its additive operands, recursively expanded through their definitions. Thus, additive operands are ignored, as they do not directly contribute to multiplicative depth; instead, their corresponding MD subtrees are used. For example, the MD tree for $c_7$ in the program $c_1 := c_2 \otimes c_3; c_4 := c_5 \otimes c_6; c_7 := c_4 \oplus c_1$, $c_4$ is as follows: $c_7$ is the root node with $c_2$, $c_3$, $c_5$ and $c_6$ as the children. Note that $c_4$ and $c_1$ aren't used; instead their MD trees are used.

The inference function, \texttt{Modswitch}, picks a leaf node in the MD tree;  inserts modulus switch operations on the operands of the node's parent; and updates the  type of the node to reflect the correct modulus level. Changes in the modulus level are then recursively propagated to all statements reached by the corresponding definition. The function \texttt{MSLevel} ensures that all operands within a statement are at a consistent modulus level.
The definition of \texttt{MSLevel} and \texttt{TypeCheck} are straightforward and are thus omitted. Note that every AST transformation uses \ila’s type system as a validation framework. Particularly, the algorithm fails if the transformed program does not type check (line $17$-$18$).

The worst-case runtime complexity of the algorithm is $O(\ell \cdot n)$, where $\ell$ is the length of the modulus chain and $n$ is the number of multiplications in the circuit. In other words, each multiplication may undergo up to $\ell$ modulus switch operations. Although achieving optimality is not a goal of this work, the algorithm aims to minimize the number of switching operations while preserving the correctness of the transformation.

The MS inference algorithm is designed for loop-free programs; therefore, any loops must be unrolled prior to the inference step.
For example, suppose the last iteration of the PSI program (Figure~\ref{fig:psi}) causes an overflow. In this case, the inference process would insert modulus switching operations into lines 14–16 of the unrolled version of the program. In practice, nested loop depth is typically less than~$5$; as such, loops can be unrolled without significantly increasing runtime complexity.

\section{Implementation} \label{sec:impl}

We extend the core language in Figure~\ref{fig:ilasyntax} with scheme-specific homomorphic operations as well as with other user-friendly features such as finite loops, vectors and matrices in a straightforward manner. Note that all loops have to terminate; otherwise the program violates Theorem~\ref{thm:msg_equiv_commands}. Under the hood, matrices are implemented as vectors. Interestingly, indexing vector might lead to out-of-bounds accesses if the index is not a constant. Since all \ila programs are terminating, a simple static analysis can help detect OOB accesses. However,  it is orthogonal to the current work.

We implement \ila  as a retargetable compiler with  support for three popular backends---SEAL, OpenFHE and TFHE-rs---for executing the FHE circuits. It supports BGV, BFV and TFHE schemes. The compiler is implemented in Python and invokes backends either by using Python bindings (OpenFHE and SEAL) or by generating Rust code (TFHE-rs). During the setup phase, it instantiates all parameters  and constructs the initial environment; statically type checks the environment and program; and if successful, invokes the corresponding backend to execute the circuit. 

There are a couple of implementation challenges.
First, FHE circuits inherently involve arbitrarily large number arithmetic (e.g., the coefficient modulus can be as large as $10^{250}$).
Fortunately, Python supports accurate arithmetic for integers. For floating point calculations, we rely on high-precision libraries such as numpy.
Second, to determine the noise overflow statically, we need a heuristic estimate to estimate noise growth. In our current implementation we employed the worst case---most conservative---noise heuristics for \ilabfv \cite{ilia2019}, \ilabgv~\cite{ana2} and average case heuristics for TFHE~\cite{tfhe}.

\section{Evaluation}\label{sec:eval}

Our evaluation aims to answer two broad questions. First, \emph{is \ila
expressive}, in the sense of handling complex FHE circuits? Second, \emph{is \ila efficient?}

%\vspace{-0.5em}
\subsection{Expressiveness}

We evaluated a series of benchmarks that varied in both multiplicative depth (up to approximately 22 levels) and circuit breadth (up to approximately 8000 multiplicative gates). The breadth of a circuit refers to the number of multiplicative gates it contains. While \ila is theoretically capable of supporting circuits with arbitrary depth and breadth, in practice, its capabilities are limited by the constraints of the underlying cryptographic libraries.

Achieving higher multiplicative depth requires a larger coefficient modulus. In Microsoft SEAL, the maximum supported bit count for the coefficient modulus is 881 bits~\cite{seal_bitcount}, which, in our experiments, corresponded to a maximum multiplicative depth of 21. OpenFHE, employs the Residue Number System (RNS) to represent the coefficient modulus. Although the use of the Residue Number System (RNS) enables support for larger coefficient moduli, it also incurs significantly higher memory overhead~\cite{fhe_rns} and necessitates larger cache sizes for efficient execution~\cite{openfhe}. In our experiments, OpenFHE programs with a multiplicative depth greater than 22 could not be executed successfully.

The breadth of a circuit is also limited by ciphertext size. Each ciphertext is represented as a degree-$d$ polynomial, making its size proportional to $d$. For instance, when $d = 32768$, storing $8000$ distinct ciphertexts would require more than $50$ GB of memory.

Our evaluation includes  all combinations of cipher-cipher and cipher-plain operations. The benchmarks also include computing functions for  exponentiation (proxy for multiplicative depth) and fibonacci (proxy for additive depth). This shows that \ila compiler can handle large FHE circuits that are both wide and deep.

Additionally,  to demonstrate that \ila can handle complex FHE circuits, we implemented a \emph{private set intersection} (PSI), \emph{private information retrieval} (PIR) and an image processing application. Each of these applications have complex control flow operations involving nested (finite) loops and use rich data structures such as vectors and matrices. PIR models database with $1000$ rows.\footnote{Similar to \emph{Wide Database} of Apple's caller ID PIR~\cite{apple-pir}.} All entries of the database are packed into a plaintext vector and the query as a ciphertext vector containing the secret index.  A cipher-plain vector multiplication yields the required  information. The image processing application uses matrix (size 1000x1000) operations to apply a filter (can be encrypted or plaintext) to the encrypted image. Note that for larger circuits, we optionally rely on an encoding that packs multiple ciphertexts reducing the size of the circuit. 

In all these cases, \ila is able to detect overflows when the multiplicative depth exceeds the noise budget. When type checking succeeds, the compiler invokes the corresponding backend to execute the circuit. Note that all applications can run on multiple backends using different schemes. For example, PIR can run on both SEAL and OpenFHE by selecting BGV/BFV scheme; PSI can run on all backends for all schemes.

\subsection{Efficiency}
A type checker that rejects all programs is trivially sound but not useful. In other words, too many false positives limit the practical use. To demonstrate the practicality of \ila, we evaluated \ila's instantiations against various measures.

Since noise estimators employed by \ila type checker are conservative, we evaluated if the type checker admits practical FHE circuits. We performed various experiments to measure how tight \ila's noise estimators are for \ilabgv and \ilabfv. This is measured using the multiplicative depth---the maximal number of sequential homomorphic multiplications that can be performed on fresh ciphertexts without noise overflow. To a good approximation, it can be used as a proxy for the lower bound of the circuit's size~\cite{accelerators, aubry}.

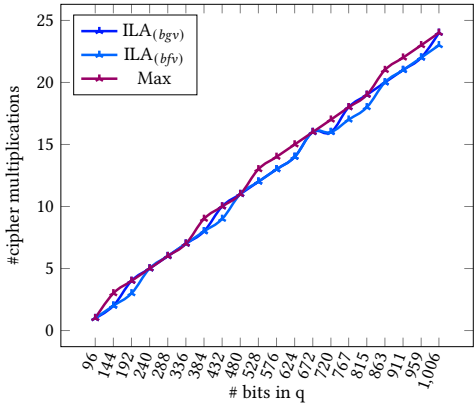
\begin{figure}
  \begin{tikzpicture}[scale=0.8]
    \begin{axis}[
    xlabel=\# bits in q,
    ylabel=\#cipher multiplications,
    xticklabel style={rotate=70,anchor=east},
    xtick=data,
    table/col sep=comma,
    legend pos= north west,
    ylabel near ticks
    ]
    \addplot[color=blue!90!cyan,smooth,mark=Mercedes star,tension=0.7,very thick] table [y=ILA_BGV, x=q]{eval/graphs/md-ila-actual.csv};
    \addlegendentry{\ilabgv}
    \addplot [color=blue!60!cyan,smooth,mark=Mercedes star,tension=0.7,very thick] table [y=ILA_BFV, x=q]{eval/graphs/md-ila-actual.csv};
    \addlegendentry{\ilabfv}
    \addplot[color=red!60!blue,smooth,mark=Mercedes star,tension=0.7,very thick] table [y=Actual_BFV, x=q]{eval/graphs/md-ila-actual.csv};
    \addlegendentry{Max}
    \end{axis}
\end{tikzpicture}
  \caption{Multiplicative depth of \ila vs possible maximum.}\label{fig:md}
\end{figure}

\begin{figure}
  \begin{tikzpicture}[scale=0.8]
    \begin{axis}[
    xmode=log,
    log basis x={2},
    ymode=log,
    xlabel=\# computations,
    ylabel=Runtime in milliseconds,
    xticklabel style={rotate=70,anchor=east},
    xtick=data,
    table/col sep=comma,
    legend pos= north west,
    ylabel near ticks
    ]
    \addplot[color=blue!50!cyan,smooth,mark=*,tension=0.7,very thick] table [y=ILA in ms, x=p]{eval/ila-tfhe.csv};
    \addlegendentry{ILA}
    \addplot[color=red!80,smooth,mark=*,tension=0.7,very thick] table [y=TFHE in ms, x=p]{eval/ila-tfhe.csv};
    \addlegendentry{TFHE-rs}
    \addplot [only marks,mark=*,mark size=1pt,
    nodes near coords={
        $\pgfmathprintnumber{\pgfkeysvalueof{/data point/y}}$
},
nodes near coords align={horizontal}, 
every node near coord/.append style={anchor=north west
},
    x filter/.code={
    \ifnum \coordindex=1
    \else
    \ifnum \coordindex=4
    \else
      \ifnum \coordindex=7
      \else
    \ifnum \coordindex=9
      \else
        \def\pgfmathresult{}
        \fi
        \fi
      \fi
     \fi                            
     }]
     table[x=p, y=TFHE in ms]{eval/ila-tfhe.csv};
     \addplot [only marks,mark=*,mark size=1pt,
    nodes near coords={
        $\pgfmathprintnumber{\pgfkeysvalueof{/data point/y}}$
},
        nodes near coords align={horizontal}, 
        every node near coord/.append style={anchor=south
        },
    x filter/.code={
    \ifnum \coordindex=1
    \else
    \ifnum \coordindex=4
    \else
      \ifnum \coordindex=7
      \else
    \ifnum \coordindex=9
      \else
        \def\pgfmathresult{}
        \fi
        \fi
      \fi
     \fi                            
     }]
     table[x=p, y=ILA in ms]{eval/ila-tfhe.csv};
    \end{axis}
\end{tikzpicture}
  \caption{\ilatfhe vs TFHE value overflow runtimes.}\label{fig:compare-tfhe}
\end{figure}
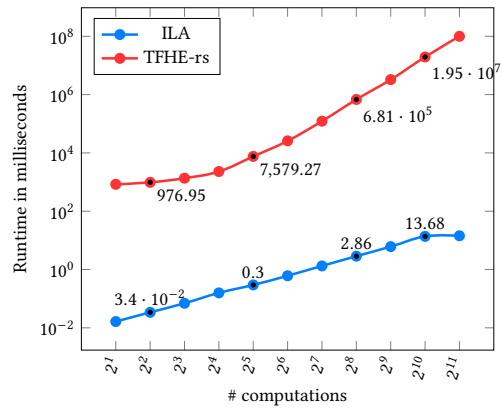

For TFHE, since noise is less of a concern, we measured the runtime of value overflow detection. Figures~\ref{fig:md} and \ref{fig:compare-tfhe} summarize the results. 

\textbf{Multiplicative depth.}  To compute the multiplicative depth for a given coefficient modulus $q$, we multiply a sequence of fresh ciphers until \ila type checker detected a noise overflow.
Using the same scheme parameters, we check whether is a true noise overflow in both SEAL and OpenFHE libraries. Note that the libraries may admit higher multiplicative depth and we record the maximal depth.
Figure~\ref{fig:md} plots  the  multiplicative depth obtained by varying the coefficient modulus $q$. It shows that  \ila's multiplicative depth  closely matches  the actual maximal multiplicative depth for both BGV and BFV schemes.

  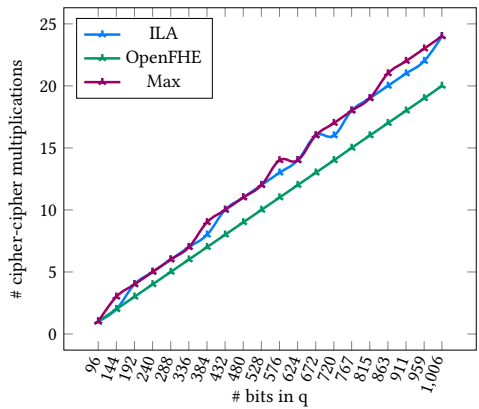
\begin{figure}
      \begin{tikzpicture}[scale=0.8]
        \begin{axis}[
        xlabel=\# bits in q,
        ylabel=\# cipher-cipher multiplications,
        xticklabel style={rotate=70,anchor=east},
        xtick=data,
        table/col sep=comma,
        legend pos= north west,
        ylabel near ticks
        ]
        \addplot[color=blue!50!cyan,smooth,mark=Mercedes star,tension=0.7,very thick] table [y=ILA, x=q]{eval/graphs/openfhe-cipher-bgv.csv};
        \addlegendentry{ILA}
        \addplot[color=green!60!blue,smooth,mark=Mercedes star,tension=0.7,very thick] table [y=OpenFHE, x=q]{eval/graphs/openfhe-cipher-bgv.csv};
        \addlegendentry{OpenFHE}
        \addplot [color=red!60!blue,smooth,mark=Mercedes star,tension=0.7,very thick] table [y=Actual, x=q]{eval/graphs/openfhe-cipher-bgv.csv};
        \addlegendentry{Max}
        \end{axis}
    \end{tikzpicture}
    
  \caption{Multiplicative depth of \ilabgv vs OpenFHE.\label{fig:compare-openfhe-bgv}}
  \end{figure}

Given that OpenFHE is a industrial strength library, we refine the above evaluation by comparing the multiplicative depth of \ila and OpenFHE. Figure~\ref{fig:compare-openfhe-bgv} shows that  \ila always outperforms OpenFHE. This demonstrates that \ila's noise overflow analysis is useful and admits practical FHE circuits.

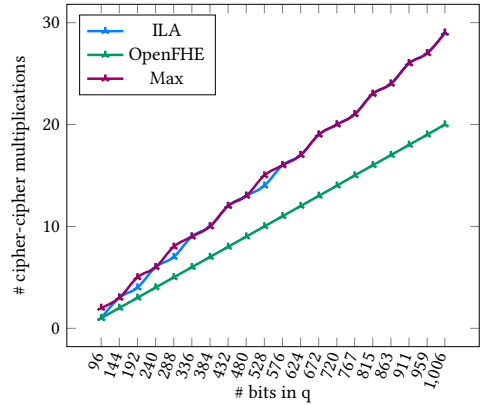
\begin{figure}
  \begin{tikzpicture}[scale=0.8]
    \begin{axis}[
    xlabel=\# bits in q,
    ylabel=\# cipher-cipher multiplications,
    xticklabel style={rotate=70,anchor=east},
    xtick=data,
    table/col sep=comma,
    legend pos= north west,
    ylabel near ticks
    ]
    \addplot[color=blue!50!cyan,smooth,mark=Mercedes star,tension=0.7,very thick] table [y=ILA, x=q]{eval/graphs/openfhe-plain-bgv.csv};
    \addlegendentry{ILA}
    \addplot[color=green!60!blue,smooth,mark=Mercedes star,tension=0.7,very thick] table [y=OpenFHE, x=q]{eval/graphs/openfhe-plain-bgv.csv};
    \addlegendentry{OpenFHE}
    \addplot [color=red!60!blue,smooth,mark=Mercedes star,tension=0.7,very thick] table [y=Actual, x=q]{eval/graphs/openfhe-plain-bgv.csv};
    \addlegendentry{Max}
    \end{axis}
\end{tikzpicture} 
  \caption{Plain-cipher mult depth of \ilabgv vs OpenFHE \label{fig:pc-compare-bgv}}
  \end{figure}

\textbf{Multiplicative depth of OpenFHE vs \ila.} Recall that OpenFHE guarantees correctness of an FHE circuit up to a preselected multiplicative depth. There are two main problems with their estimates, their estimates are not tight, and they treat cipher-cipher and plain-cipher multiplications as the same. Note that plaintext does not carry any noise. As a result plain-cipher multiplication result has less noise than a cipher-cipher multiplication. Hence, more  plain-cipher multiplications are possible than OpenFHE guarantees. In this experiment, we computed plain-cipher multiplications for BGV.  Figure~\ref{fig:pc-compare-bgv} shows that \ila over performed OpenFHE for every value of $q$. Appendix~\ref{app:eval} shows full evaluation of \ila for BGV and BFV schemes.

\textbf{Value overflow.} In the case of TFHE, the multiplication operation resets the noise in the result. Moreover, the noise growth is very slow with other homomorphic operations. Hence, runtime performance of value overflow detection is a greater concern.

Our goal is to compare the runtime performance of value overflow detections in  TFHE and \ilatfhe. To this end, we construct a FHE circuit with $t = 2^p$ additions. Note that this allows $t-1$ valid additions and  overflow occurs with the last addition.
The circuit is implemented in both \ila and TFHE-rs. For the latter, we use \emph{overflow\_add} that tracks the value overflows during the circuit execution.
We measure the time taken to detect overflows for the same number of operations in both \ila ad TFHE-rs.

Figure~\ref{fig:compare-tfhe} shows the runtime comparison of \ilatfhe and TFHE-rs for detecting value overflows when run on a Intel i$9$ 2.3 GHz 8-core processor with $16$ GB RAM.  Note that we use the average (1000 runs) execution time for \ilatfhe type checker to offset any randomness in the runs.

Evaluation  shows that TFHE-rs is orders of magnitude slower than \ilatfhe.
In all instances, \ilatfhe was able to detect value overflows in milliseconds, whereas TFHE-rs took hours for larger circuits ($p \ge 11$). Even though \ilatfhe's runtime increased with more number of computations, the rate of increase is much slower than that of TFHE-rs. This is unsurprising as FHE circuits are known to be slow, and any dynamic detection inherits the poor performance. Though a GPU support might accelerate the performance of TFHE-rs, \ilatfhe may still outperform as it does not execute the FHE circuit for overflow detection.

\textbf{Modulus Switch (MS) inference.} Recall that the MS inference algorithm introduced in Section \ref{sec:msinfer} is an optimization designed to further tighten the multiplicative depth guarantees provided by ILA. We implemented this algorithm using the Microsoft SEAL library as the backend and evaluated it across multiple modulus chains, sorted in decreasing order.

Experimental results show that the multiplicative depth inferred by the  algorithm matches the maximum achievable depth for the corresponding modulus chain. Figure~\ref{appfig:msmd} from Appendix~\ref{app:eval} illustrates the multiplicative depth before and after the MS inference.

\vspace{-0.5em}
\section{Related Work}

\textbf{Existing FHE Solutions.}
Table~\ref{tab:fhesupport} summarizes the support for correctness reasoning
among the state-of-the-art FHE libraries.
%% \footnote{Note that the existing FHE
%% compilers are typically built on the top of these libraries and thus the
%% compiler's ability to detect correctness issues is often reduced to the ability
%% of the underlying FHE library.}
Only a few libraries support more than one FHE
scheme and all of them have limited---if any---error detection support. Neither
OpenFHE~\cite{openfhe}, SEAL~\cite{sealcrypto}, Lattigo~\cite{lattigo} or
HElib~\cite{helib}  handle value overflows. As a concrete example, if the plaintext modulus is $100$, and the decrypted output is $20$, then the intended result could be one of $120$, $-20$ or $50720$.

\begin{table}
  \scalebox{1}{
\begin{tabular}{c | c | c | c | c | c }
\hline
FHE Library & Schemes & $O_{value}$ &  $O_{noise}$ &    SK Ind & Static  \\
\hline
OpenFHE~\cite{openfhe} & \color{green}\ding{51}  & \color{red}\ding{55} & \color{red} \ding{55} & \color{red}\ding{55}  & \color{red}\ding{55} \\
SEAL~\cite{sealcrypto} & \color{red}\ding{55} & \color{red}\ding{55} & \color{red}\ding{55}  & \color{red}\ding{55} &  \color{red}\ding{55} \\
Lattigo~\cite{lattigo} & \color{green}\ding{51} & \color{red}\ding{55} & \color{red}\ding{55}  & \color{red}\ding{55} &  \color{red}\ding{55} \\
HElib~\cite{helib} & \color{red}\ding{55} & \color{red}\ding{55} &\color{green}\ding{51}  & \color{green}\ding{51} & \color{red}\ding{55} \\
TFHE-rs~\cite{tfhe-rs} & \color{red}\ding{55} & \color{green}\ding{51} &\color{red}\ding{55}  & \color{green}\ding{51} & \color{red}\ding{55} \\
\hline
\ila (this work) &  \color{green}\ding{51} & \color{green}\ding{51} & \color{green}\ding{51}  &  \color{green}\ding{51} & \color{green}\ding{51} \\
\hline
\end{tabular}
}
\caption{The first column indicates support for multiple schemes. The next two columns check for value and noise overflow errors. The fourth column verifies secret key independence, and the final column indicates whether the technique is static.\label{tab:fhesupport}}
\end{table}

SEAL, Lattigo, HElib and OpenFHE~\cite{openfhe} support BGV and BFV schemes including modulus switching operation. None of them support modulus switch inference. %With the exception of HElib none of them offer noise overflow detection.
OpenFHE and TFHE-rs~\cite{tfhe-rs} also support TFHE although the former's support is limited in functionality.
%Both the libraries convert a function to a lookup table (LUT) and apply it to the ciphertext.
%Since the function may exhaust all noise, they perform a PBS to invoke the LUT.
%In most cases, it is not possible to disable PBS as they do not have any noise overflow detection. Thus, a circuit may use PBS even when not required.

%All libraries, except TFHE-rs, also support the approximate CKKS scheme.

%\AG{Explain the importance of measuring multiplicative depth}

Noise overflow issues differ with every library. OpenFHE guarantees are valid up to a preselected depth (chosen by the user) of cipher-cipher operations; however, the guarantees get invalidated in the presence of other homomorphic operations (e.g., cipher-plain multiplication).  To overcome this limitation, OpenFHE recommends estimating the depth parameter that over-approximates all operations as cipher-cipher~\cite{openfhe_md}. Since noise growth is slower for other combinations, the over-approximation leads to a much slower circuit (due to larger parameters). Moreover, it is tricky to get these estimations correct as the depth is a function of fresh ciphertext operations whereas the programs often operate on a series of mutated ciphertexts. %For example, the estimated depth for $ct_1 = ct_1 * ct_1; ct_1 = ct_1 * ct_1$ is $3$ and not $2$~\cite{openfhe_correct}.

SEAL and Lattigo do not offer any noise estimation whereas HElib supports explicit user queries on noise estimate queries at runtime. Note that the latter setup is undesirable for two reasons. First, a user has to explicitly query whether an overflow has occurred. Second, it can be orders of magnitude slower as  FHE circuits are well-known to suffer from poor performance.

TFHE-rs is more nuanced: it supports detecting value overflows but not  noise overflows. However, value overflow detection requires running the FHE circuit which is once again slow. In this work, we use a static approach that can be orders of magnitude fast (Section~\ref{sec:eval}). Interestingly, it has a noise management strategy that automatically inserts PBS operations after a preselected operation depth (controlled by the user). Hence, noise overflow might appear to be of lesser concern. However, we argue that it isn't. First, the user may choose to turn it off for some operations (e.g., addition). Second, the operation depth for PBS  is chosen apriori and is agnostic to whether a PBS is required. As PBS is computationally expensive, this approach is sub-optimal and further slows down the circuit. More generally, precise PBS insertion  would entail noise overflow detection that TFHE-rs lacks.

%Only TFHE-rs supports value overflow dectection. TFHE-rs uses circuit simulation to detect value overflows.
%However, circuit simulation is computationally expensive.

\textbf{Other FHE Compilers.}
We refer the reader to a survey on the state-of-the-art in FHE compilers~\cite{sok}. To the best of our knowledge, \ila is the only work to prove both functional correctness property.
\textsc{Alchemy}~\cite{alchemy} is an embedded DSL in Haskell for
using FHE with type-level tracking of ciphertext noise; however, neither do
these type-level noise computations come with formal guarantees, nor are the
heuristics for noise estimations sound.

Other FHE compilers such as EVA~\cite{eva}, Coyote~\cite{coyote} and others~\cite{chielle2018e3,alchemy,sok} focus on efficient encoding schemes but not formal guarantees. They are  orthogonal to our work.

\vspace{-0.5em}
\section{Conclusion and Future Work}
We present \ila, a correctness-oriented IR and an abstract model for FHE circuits.
Our IR is backed by a type system that statically tracks low-level quantitative bounds (e.g., ciphertext noise) without using the secret key. Using our type system, we identify and prove a strong \emph{functional correctness} criterion for \ila circuits. Furthermore, we instantiate \ila with three absolute schemes---BGV, BFV and TFHE---and get functional correctness for free. Comparative evaluation of \ila against three popular FHE libraries shows that \ila's static analysis is sound, tight and efficient.

In future work, we aim to extend \ila to support functional, higher-order programming with
FHE. While function types are straightforward, 
they are not ergonomic without 
\emph{bound polymorphism}, which would enable types such as $\forall
\alpha.\ \mathsf{cipher}\ \alpha \to \mathsf{cipher}\ (2 * \alpha)$. 
Bound polymorphism in our setting requires extending the noise inference
algorithms of the underlying cryptosystems (e.g., BGV) to handle \emph{symbolic}
bounds, which will likely require insights from nonlinear arithmetic constraint
solving.

Additionally, we aim to instantiate \ila with approximate FHE schemes such as
CKKS~\cite{ckks}. This extension primarily requires our notion of an \ila model
to be generalized so that Commutativity does not hold on the nose, but holds
with some bounded amount of error.

\begin{acks}
This work was supported in part by the National Science Foundation under
Grant No.\ NSF 2348304 (Anitha Gollamudi) and Grant No.\ NSF 2224279 (Joshua Gancher).
\end{acks}

\bibliographystyle{plainnat}
%\bibliography{ref}

\clearpage
\appendix
\pagebreak

\section{Modulus Switching}\label{app:ms}

Since the focus of  modulus switch inference is to demonstrate the role of the type system in transformation validation, we keep it simple. Our inference algorithm  works for loop-free programs.  Multiplicative depth computation for ciphertexts inside a loop  depends on the loop iteration. One potential way to fix this would be to carry out the inference after  unrolling the loop. Since \ila programs only have finite loops, this is feasible.

As an example, suppose that the last iteration of the PSI program (Figure~\ref{fig:psi}) overflows. Then this approach would yield a program where  lines $14$-$16$ in the unrolled version of the program have modulus switching operations as shown below.
\begin{lstlisting}[language=Python, style=mystyle, mathescape=true, firstnumber=12]
...              # unroll the inner loop
t5 = A[0]        # last iteration
t6 = modswitch(t5) $\oplus$ modswitch(t2) # switch levels  
t7 = modswitch(result) $\otimes$ t6    # level propagation
result = t7 $\otimes$ modswitch(t4)    
\end{lstlisting}

\section{Evaluation}\label{app:eval}

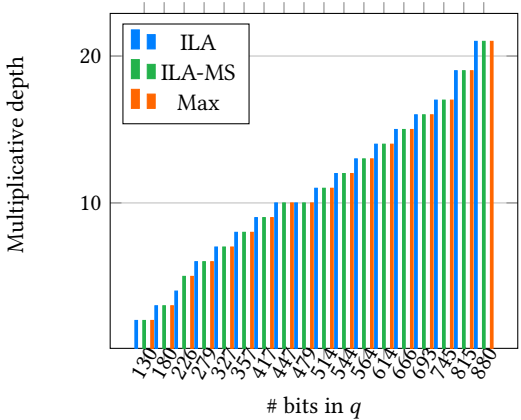
\begin{figure}
    \begin{tikzpicture}
    \begin{axis}[width=7cm,
        ybar,
        bar width=1pt,
        xtick=data,
        table/col sep=comma,
        xticklabels from table={eval/graphs/seal-ms-latest.csv}{Q},
        xlabel={\# bits in $q$},
        ylabel = {Multiplicative depth},
        xticklabel style={yshift=10pt, rotate=60},
        ylabel style = {xshift=10pt},
        ymajorgrids,
        legend pos=north west]
        \addplot[fill=blue!50!cyan,draw=blue!50!cyan] table [x expr=\coordindex, y=ILA]{eval/graphs/seal-ms-latest.csv};
        \addplot[fill=yellow!20!green,draw=yellow!20!green] table [x expr=\coordindex, y=ILA-MS]{eval/graphs/seal-ms-latest.csv};
        \addplot[fill=orange!80!red,draw=orange!80!red] table [x expr=\coordindex, y=Actual]{eval/graphs/seal-ms-latest.csv};
        \legend{ILA, ILA-MS, Max}
    \end{axis}
\end{tikzpicture}
    \caption{Multiplicative depth before and after modulus switch inference.}\label{appfig:msmd}
  \end{figure}

\begin{figure}
      \begin{tikzpicture}[scale=0.8]
        \begin{axis}[
        xlabel=\# bits in q,
        ylabel=\# cipher-cipher multiplications,
        xticklabel style={rotate=70,anchor=east},
        xtick=data,
        table/col sep=comma,
        legend pos= north west,
        ylabel near ticks
        ]
        \addplot[color=blue!50!cyan,smooth,mark=Mercedes star,tension=0.7,very thick] table [y=ILA, x=q]{eval/graphs/openfhe-cipher-bfv.csv};
        \addlegendentry{ILA}
        \addplot[color=green!60!blue,smooth,mark=Mercedes star,tension=0.7,very thick] table [y=OpenFHE, x=q]{eval/graphs/openfhe-cipher-bfv.csv};
        \addlegendentry{OpenFHE}
        \addplot [color=red!60!blue,smooth,mark=Mercedes star,tension=0.7,very thick] table [y=Actual, x=q]{eval/graphs/openfhe-cipher-bfv.csv};
        \addlegendentry{Max}
        \end{axis}
    \end{tikzpicture}
  \caption{Multiplicative depth of \ilabfv vs OpenFHE. X-axis shows the number of bits in coefficient modulus q; Y-axis shows number of plain-cipher multiplications estimated by ILA, OpenFHE vs maximal possible (higher is better).\label{appfig:compare-openfhe-bfv}}
  
  \end{figure}
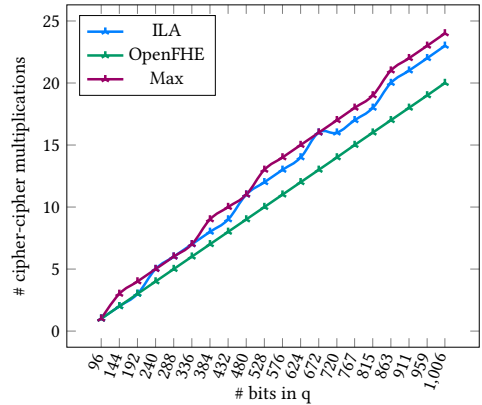

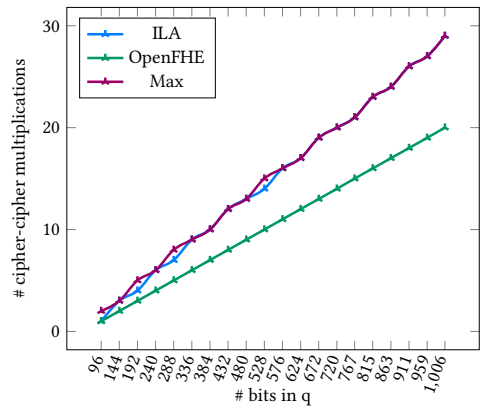
\begin{figure}
  \begin{tikzpicture}[scale=0.8]
    \begin{axis}[
    xlabel=\# bits in q,
    ylabel=\# cipher-cipher multiplications,
    xticklabel style={rotate=70,anchor=east},
    xtick=data,
    table/col sep=comma,
    legend pos= north west,
    ylabel near ticks
    ]
    \addplot[color=blue!50!cyan,smooth,mark=Mercedes star,tension=0.7,very thick] table [y=ILA, x=q]{eval/graphs/openfhe-plain-bgv.csv};
    \addlegendentry{ILA}
    \addplot[color=green!60!blue,smooth,mark=Mercedes star,tension=0.7,very thick] table [y=OpenFHE, x=q]{eval/graphs/openfhe-plain-bgv.csv};
    \addlegendentry{OpenFHE}
    \addplot [color=red!60!blue,smooth,mark=Mercedes star,tension=0.7,very thick] table [y=Actual, x=q]{eval/graphs/openfhe-plain-bgv.csv};
    \addlegendentry{Max}
    \end{axis}
\end{tikzpicture} 
  \caption{Plain-cipher multiplicative depth of \ilabgv vs OpenFHE.\label{appfig:pc-compare-bgv}}
  \end{figure}

Evaluation shows comparative multiplicative depth of \ila against SEAL and OpenFHE for BGV and BFV schemes.
Figures \ref{appfig:compare-openfhe-bgv}, \ref{appfig:compare-openfhe-bfv}, \ref{appfig:pc-compare-bgv} and \ref{appfig:pc-compare-bfv} compare \ila with OpenFHE  for cipher-cipher and cipher-plain multiplications.
Additionally, Figure~\ref{appfig:msmd} compares the multiplicative depth before and after modulus switching inference.

  \begin{figure}
      \begin{tikzpicture}[scale=0.8]
        \begin{axis}[
        xlabel=\# bits in q,
        ylabel=\# cipher-cipher multiplications,
        xticklabel style={rotate=70,anchor=east},
        xtick=data,
        table/col sep=comma,
        legend pos= north west,
        ylabel near ticks
        ]
        \addplot[color=blue!50!cyan,smooth,mark=Mercedes star,tension=0.7,very thick] table [y=ILA, x=q]{eval/graphs/openfhe-cipher-bgv.csv};
        \addlegendentry{ILA}
        \addplot[color=green!60!blue,smooth,mark=Mercedes star,tension=0.7,very thick] table [y=OpenFHE, x=q]{eval/graphs/openfhe-cipher-bgv.csv};
        \addlegendentry{OpenFHE}
        \addplot [color=red!60!blue,smooth,mark=Mercedes star,tension=0.7,very thick] table [y=Actual, x=q]{eval/graphs/openfhe-cipher-bgv.csv};
        \addlegendentry{Max}
        \end{axis}
    \end{tikzpicture}
    
  \caption{Multiplicative depth of \ilabgv vs OpenFHE. X-axis shows the number of bits in coefficient modulus q; Y-axis shows number of plain-cipher multiplications estimated by ILA, OpenFHE vs maximal possible (higher is better).\label{appfig:compare-openfhe-bgv}}
  \end{figure}
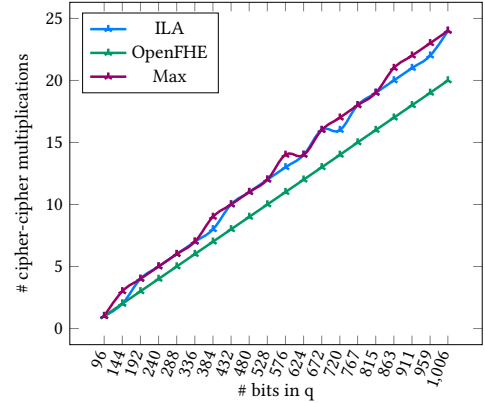
  
\begin{figure}
  \begin{tikzpicture}[scale=0.8]
    \begin{axis}[
    xlabel=\# bits in q,
    ylabel=\# cipher-cipher multiplications,
    xticklabel style={rotate=70,anchor=east},
    xtick=data,
    table/col sep=comma,
    legend pos= north west,
    ylabel near ticks
    ]
    \addplot[color=blue!50!cyan,smooth,mark=Mercedes star,tension=0.7,very thick] table [y=ILA, x=q]{eval/graphs/openfhe-plain-bfv.csv};
    \addlegendentry{ILA}
    \addplot[color=green!60!blue,smooth,mark=Mercedes star,tension=0.7,very thick] table [y=OpenFHE, x=q]{eval/graphs/openfhe-plain-bfv.csv};
    \addlegendentry{OpenFHE}
    \addplot [color=red!60!blue,smooth,mark=Mercedes star,tension=0.7,very thick] table [y=Actual, x=q]{eval/graphs/openfhe-plain-bfv.csv};
    \addlegendentry{Max}
    \end{axis}
\end{tikzpicture}
  \caption{Plain-cipher multiplicative depth of \ilabfv vs OpenFHE. X-axis shows the number of bits in coefficient modulus q; Y-axis shows number of plain-cipher multiplications estimated by ILA, OpenFHE vs maximal possible (higher is better).}\label{appfig:pc-compare-bfv}
\end{figure}

\section{Proofs}\label{app:proofs}

\begin{theorem}[Semantic Safety: Expressions] \label{thm:semsafety-exp}
Suppose that $\topub(\sparam); \Gamma \vdash e : \tau$. Then $\sparam; \Gamma \vDash e : \tau$.
\end{theorem}
\begin{proof}
    Recall that we must show that, for all $\gamma$ such that $\sparam; \Gamma
    \vDash \gamma$ (i.e., $\gamma(x) \in \sem{\Gamma(x)}^\sparam$ for all $x \in
    \Gamma$), there exists a $v$ such that $\langle \gamma, e \rangle \Downarrow
    v$ and $v \in \sem{\tau}^\sparam$. 

    By induction on the typing derivation (Figure~\ref{fig:ilaexptr}). 
    \begin{itemize}
        \item Case \textsc{var}: We have that $\langle \gamma, x
            \rangle \Downarrow \gamma(x)$, $\gamma(x) \in
            \sem{\Gamma(x)}^\sparam$, and $\tau = \Gamma(x)$. The result
            follows.
        \item Case \textsc{sub}: The result follows once we show that, for all
            sorts $s \in \{\Cipher, \Plain, \msg\}$, if $\alpha \leq_s \alpha'$,
            then $\sem{s\ \alpha}^\sparam \subseteq \sem{s\ \alpha'}$. Since
            $\sem{s\ \alpha}^\sparam = \{v \in \sem{s} \mid |v|_s^\sparam \leq_s
            \alpha\}$, this follows from transitivity of $\leq_s$.
        \item Case \textsc{Const}: We need to show that, for $s \in \{\msg,
            \Plain\}$, for all 
                $v \in \sem{s}$, $v \in \sem{s\ |v|_s^{\topub(\sparam)}}^\sparam$. 
                Hence, we must show that $|v|_s^{\topub(\sparam)} \leq_s
                |v|_s^\sparam$. For $s \in \{\msg, \Plain\}$, the right bound is
                defined to be the left. 
        \item Case \textsc{op}: We know that for all $i$, $\sparam; \Gamma
            \vDash e_i : \tau_i$ and that
            $\sem{op}_\mathsf{bnd}^{\topub(\sparam)}(\overrightarrow{|\tau_i|}) = \alpha$,
            and we must prove that $\sparam; \Gamma \vDash \op(\overrightarrow{e_i}) : s\
            \alpha$. 

            To this end, take $\gamma$ such that $\sparam; \Gamma \vDash
            \gamma$.
            By induction, we have that $\langle \gamma, e_i \rangle \Downarrow
            v_i \in \sem{\tau_i}^\sparam$ for all $i$, 
            and thus $\langle \gamma,
            \op(\overrightarrow{e_i}) \rangle \Downarrow v = \sem{\op}(\overrightarrow{v_i})$. 
            We know that $\sort(v) = s$.
            We must show that $|\sem{\op}(\overrightarrow{v_i})|_s^\sparam \leq
            \alpha$. 
            First, observe that
            $\sem{\op}_\bnd^{\topub(\sparam)}(\overrightarrow{|v_i|_{\sort(\tau_i)}}^\sparam)$ $\leq$ 
              $\sem{\op}_\bnd^{\topub(\sparam)}(\overrightarrow{|\tau_i|})$ $=\alpha$
            by the Downwards Closed axiom.
            Then, by the Commutativity axiom, we have
            $|\sem{\op}(\overrightarrow{v_i})_s^\sparam| \leq_s \alpha.$ 
    \end{itemize}
\end{proof}

\begin{theorem}[Semantic Safety: Commands]
    Suppose that $\topub(\sparam); \Gamma \vdash c \dashv \Gamma'$. Then
    $\sparam; \Gamma \vDash c \Dashv
    \Gamma'$.
\end{theorem}
\begin{proof}
    Recall that we need to show that whenever $\sparam; \Gamma \vDash \gamma$
    and $\langle \gamma, c \rangle \Downarrow \gamma'$, that $\sparam; \Gamma'
    \vDash \gamma'$. 

    Induct on the typing derivation of $c$ (Figure~\ref{appfig:ilacmdtr}):
    \begin{itemize}
        \item Case \textsc{skip}: immediate, since $\langle \gamma,
            \mathsf{skip}
            \rangle \Downarrow \gamma$.
        \item Case \textsc{Assgn}: we have that $\langle \gamma, x := e \rangle
            \Downarrow \gamma[x \mapsto v]$, where $\langle \gamma, e \rangle
            \Downarrow v$, and we must show that $\sparam; \Gamma[x \mapsto
            \tau] \vDash \gamma[x \mapsto v]$. 
            By Theorem~\ref{thm:semsafety-exp}, we have that $v \in \sem{\tau}
            \in \sem{\tau}^\sparam$. 
            Since we are updating $\Gamma$ to always point to $\tau$, the result
            follows. 
        \item Case \textsc{seq}: 
            immediate, since we know by induction that any post-state of $c_1$
            will satisfy its output type context, which is assumed to be the
            input type context of $c_2$.
        \item Case \textsc{if}: we have that $\langle \gamma,
            \eif{e}{c_1}{c_2}\rangle$ evaluates as $\langle \gamma, c_1 \rangle
            \Downarrow \gamma_1$
            if $\langle \gamma, e \rangle$ evaluates to $\sem{true}()$, and
            $\langle \gamma, c_2 \rangle \Downarrow \gamma_2$ otherwise.
            By induction, we have that $\sparam; \Gamma_1 \vDash \gamma_1$ and
            $\sparam; \Gamma_2 \vDash \gamma_2$. 
            We must show that if $\Gamma_1 \sqcap \Gamma_2 \sim \Gamma'$, then 
            $\sparam; \Gamma' \vDash \gamma_1$ and 
            $\sparam; \Gamma' \vDash \gamma_2$.

            Take $x \in \Gamma'$. Then, by assumption, $x \in \Gamma_1$ and
            $\Gamma_1(x) \leq \Gamma'(x)$. Since $\gamma_1(x) \in
            \sem{\Gamma_1}^\sparam$, we have that $\gamma_1(x) \in
            \sem{\Gamma'}^\sparam$ since subtyping respects
            $\sem{\cdot}^\sparam$. The argument is the same for $\Gamma_2$.
    \end{itemize}
\end{proof}

\begin{theorem}[Message Equivalence: Expressions]\label{thm:msgequiv-exp}
    Suppose that $\topub(\sparam); \Gamma \vdash e : \tau$ and $\sparam; \Gamma \vDash \gamma$. 
    Let \[\gamma' = \{ x \mapsto \interp^\sparam_{\sort(\Gamma(x))}(\gamma(x))
    \}.\]
Then,
$\config{\gamma, e} \Downarrow v$ and $\config{\gamma', e}
    \Downarrow^\sparam_\msg v_\msg$ such that
    $\interp^\sparam_{\mathsf{sort}(v)}(v) = v_\msg.$
\end{theorem}
\begin{proof}
By induction on the typing derivation. 
    \begin{itemize}
        \item Case \textsc{var}: we have that if $e = x$, then $v=\gamma(x)$ and 
            $v_\msg = \gamma'(x) =
            \interp^\sparam_{\sort(\Gamma(x))}(\gamma(x))$. The result follows
            from the fact that $\sort(\gamma(x)) = \sort(\Gamma(x))$, since $\sparam;
            \Gamma \vDash \gamma$.
        \item Case \textsc{sub}: the inductive hypothesis on $e$ is equivalent
            to the result, since the sort of $\tau$ does not change (only the
            bound). 
        \item Case \textsc{const}: if $e = v \in \sem{s}$, $\langle \gamma, v \rangle
            \Downarrow v$ and $\langle \gamma', v \rangle
            \Downarrow^\sparam_\msg \interp^\sparam_s(v)$. The result follows. 
        \item Case \textsc{op}: 
            We have that 
            $e = \op(\overrightarrow{e_i})$ where 
            $\op : \overrightarrow{s_i} \to s$,
            and $\sem{op}_\bnd^{\topub(\sparam)}(\overrightarrow{|\tau_i|})$ is
            defined. 
            Inductively, we have that for each $i$, $\langle \gamma, e_i \rangle \Downarrow
            v_i$ and $\langle \gamma', e_i \rangle \Downarrow^\sparam_\msg v_{i,
            \msg}$ such that 
                $\interp^\sparam_{s_i}(v_i) = v_{i, \msg}.$
            We also have that
            $\langle \gamma, \op(\overrightarrow{e_i}) \rangle
            \Downarrow$ $\sem{op}(\overrightarrow{v_i})$ and that 
            $\langle \gamma', \op(\overrightarrow{e_i}) \rangle
            \Downarrow^\sparam_\msg$ $\sem{op}_\msg(\overrightarrow{v_{i,\msg}})$. 
            Also, observe that $\sem{\op}(\overrightarrow{v_i}) \in \sem{s}$ by
            assumption. 

            Putting everything together, we need to show that 
            \[ 
                \interp^\sparam_s(\sem{op}(\overrightarrow{v_i}))
                =
                \sem{\op}_\msg(\overrightarrow{\interp^\sparam_{s_i}(v_i)}).
                \]
            This is the conclusion of Commutativity in the model. 

            To apply Commutativity, we must show that 
            \[\sem{\op}_\bnd^{\topub(\sparam)}(\overrightarrow{|v_i|^\sparam_{s_i}})\]
            is defined.
            From Semantic Safety, we have that 
            $v_i \in \sem{\tau_i}^\sparam$ for all $i$, and hence 
            $|v_i|^\sparam_{s_i} \leq_{s_i} |\tau_i|$.
            Since $\sem{op}_\bnd^{\topub(\sparam)}(\overrightarrow{|\tau_i|})$
            is defined, the result follows from the Downwards Closed axiom in
            the model.
    \end{itemize}
\end{proof}

\begin{theorem}[Message Equivalence: Commands] 
    Suppose that $\topub(\sparam); \Gamma \vdash c \dashv \Gamma'$ and $\sparam; \Gamma \vDash \gamma$.
Let $\gamma' = \{ x \mapsto \interp^\sparam_{\sort(\Gamma(x))}(\gamma(x))
\}.$ Then, if 
$\config{\gamma, c} \Downarrow \gamma_1$, we have that 
$\config{\gamma', c} \Downarrow^\sparam_\msg \gamma_1'$ such that, for all $x \in
    \gamma_1$, $\interp^\sparam_{\mathsf{sort}(\gamma_1(x))}(
\gamma_1(x)) = \gamma_1'(x)$ .
\end{theorem}
\begin{proof}
    Follows by induction on the typing derivation.
    \begin{itemize}
        \item Case \textsc{skip}: We have that $\langle \gamma, \mathsf{skip}
            \rangle \Downarrow \gamma$ and $\langle \gamma', \mathsf{skip}
            \rangle \Downarrow^\sparam_\msg \gamma'$. The result follows by
            assumption.
        \item Case \textsc{Assgn}: 
            We have that $\langle \gamma, y := e \rangle \Downarrow \gamma[y
            \mapsto v]$ where $\langle \gamma, e \rangle \Downarrow v$, and 
            $\langle \gamma', y := e \rangle \Downarrow_\msg^\sparam \gamma'[y
            \mapsto v']$ where $\langle \gamma', e \rangle
            \Downarrow^\sparam_\msg v'$. 

            If $x \neq y$, the result follows by assumption. If $x = y$, the
            result follows by Message Equivalence on expressions
            (Theorem~\ref{thm:msgequiv-exp}).
        \item Case \textsc{seq}: The result follows, since the postcondition of
            the inductive hypothesis for $c_1$ matches the precondition of the
            inductive hypothesis for $c_2$.
        \item Case \textsc{if}: Note that since the guard $e$ of the if
            statement has type $\msg\ \alpha$, and $\interp^\sparam_\msg(v) = v$
            for all $v$, we will have that $\eif{e}{c_1}{c_2}$ will evaluate in
            the same direction for both $\gamma$ and $\gamma'$. 
            If $e$ evaluates to $\sem{true}()$, then the result follows by the
            inductive hypothesis on $c_1$; otherwise, the result follows by the
            inductive hypothesis on $c_2$.
    \end{itemize}
\end{proof}

\section{\texorpdfstring{\ilatfhe}{} instantiation}\label{app:tfhe}
We introduce \ilatfhe, an instantiation of \ila model for the CGGI/TFHE scheme.
%The set $\spset$ contains all scheme parameters including all keys. The public parameter set $\ppset \subseteq \spset$ contains all public parameters.
 TFHE is a fast bootstrapping GLWE scheme.
%That means, plaintext polynomials are elements of rings $\mb_{t} \cup \ring{t}$. Similarly, ciphertext polynomials are elements of the set $\mb_{q}^{n+1} \cup \ring{q}^{n+1}$. 
%% We define $B_{\PSort}$  as a triple $\alpha = (id, \inf, \sup, \eps)$. Since a plaintext polynomial can be an element of either $\mb_{t}$ or $\ring{t}$, the identifier $id$ keeps track of this information. $B_{\CSort}$ as a triple $(id, \inf, \sup, \eps)$ with $ id, \inf, \sup$ inherited from the underlying plaintext. $\eps$ is the noise in the ciphertext.  We say $(id, \inf, \sup, \eps) \leq_{\CSort} (id', \inf', \sup', \eps')$ if $id = id'$ and $\inf' \le \inf \le \sup \le \sup'$ and $\eps \le \eps'$.
%% We are now ready to instantiate \ila with the TFHE scheme.

Definition~\ref{def:ilatfhe-app} shows the  full instantiation. Note that it closely follows the definition of \ila model (see Definition~\ref{def:ilamodel}, Section~\ref{sec:ila}).

\begin{definition}[\ilatfhe]\label{def:ilatfhe-app}
  The \ilabgv model is defined as follows:
  \begin{itemize}
  \item Bounds sets:
   \begin{itemize}
   \item $B_{\MSort} = \emptyset$;
    \item $B_{\PSort} = \{(\inf, \sup, \eps) \mid \inf \le \sup \text{ and } \eps \in \mathbb{N} \}$ 
      ordered by $\leq_{\PSort}$;
      \item $B_{\CSort} = \{(\id, \inf, \sup, \eps) \mid \inf \le \sup, \\ \id \in \{ \lwe, \rlwe, \gsw \} \text{ and } \eps\in \mathbb{N} \}$ 
      ordered by $(\id, \inf_1, \sup_1, \eps_1) \leq_{\CSort}  (\id, \inf_2, \sup_2, \eps_2)$ if $\inf_2 \le \inf_1 \le \sup_1 \le sup_2$ and $\eps_1 \le \eps_2$;
      \end{itemize}
  \item Mappings for bounds computation:
  \begin{itemize}
    \item $|m|^\pparam_\MSort = \emptyset $;
    \item $|p|^\pparam_\PSort = (\inf, \sup) $; computed as in \ilabgv
  \item  $|\tht|^\sparam_\CSort  = (\id, \inf, \sup, \eps)$ computation similar to \ilabgv
  \end{itemize}
  \item Decoding on plaintexts: $\interp^\pparam_\PSort(p) = \code{decode}(p)$;
  \item Decryption on ciphertexts: $\interp^\sparam_\CSort(\tht) = \code{decode}(\code{decrypt}(\tht))$;
  \item Operator $\op \in \{ \oplus, \otimes, \times, \code{cmux}, \boxtimes,  \boxdot, \code{pbs} \}$ defined as follows:
  \begin{itemize}
   \item $\sem{\otimes}$: $\{\CSort, \PSort\}$, $\{\CSort, \PSort \}$ $\to$ $\{\CSort, \PSort \}$;
   \item $\sem{\oplus}$: $\{\CSort, \PSort\}$, $\{\CSort, \PSort \}$ $\to$ $\{\CSort, \PSort \}$;
   \item $\sem{\code{cmux}}: \{\CSort, \CSort, \CSort \} \to \{\CSort \}$;
   \item $\sem{\boxdot}: \{\CSort, \CSort\} \to \{\CSort \}$;
   \item $\sem{\code{pbs}}: \{\CSort, \CSort, \CSort \} \to \{\CSort \}$;
   \item $\sem{\times}$: $\integer, \CSort$ $\to$ $\CSort$; and
   \item bounds manipulation mapping $\sem{\op}^\pparam_\bnd$ shown in Table~\ref{tab:tfhe_operations-app}.
   \end{itemize}
  \end{itemize}
  \end{definition}

We prefix 'cipher' with $\id$ whenever $\id$ is significant. For simplicity, we assume that $\lwe < \rlwe < \gsw$.
\begin{table*}

\begin{small}
    \begin{tabular}{@{}llll@{}}
        \toprule
        \multicolumn{1}{@{}c}{\textbf{Op}} &
    %    \multicolumn{2}{@{}c}{\textbf{Signature }} &
        \multicolumn{1}{c}{$\sem{\cdot}^\pparam_\bnd $} &
        \textbf{Conditions}\\[2pt]
        \bottomrule
        %%%
         $\sem{\oplus}^\pparam_\bnd $($\overrightarrow{\id_i, {\inf}_i, {\sup}_i, \eps_i}$) & =~ $(\id, \inf, \sup, \eps)$ & $\id_i \in \{ \lwe, \rlwe \}$ and $\id = \max\{ \lwe, \rlwe\}$ \\
         $\sem{\oplus}^\pparam_\bnd $($\overrightarrow{{\inf}_i, {\sup}_i}$)   &=~ $(\inf, \sup)$ & \\
         $\sem{\oplus}^\pparam_\bnd (\inf_1, \sup_1), (\id, \inf_2, \sup_2, \eps)$ &=~ $(\id, \inf, \sup, \eps)$ & \\
        
         where && \qquad and in all cases \\
         $\inf := \inf_1 + \inf_2$, $\sup := \sup_1 + \sup_2$, $\eps := \eps_1 + \eps_2$ && $-t/2 \le \inf \le \sup < t/2$ and $\eps \le q/2t$ \\
        \hline
        $\sem{\otimes}^\pparam_\bnd $($\overrightarrow{\inf_i, \sup_i}$)  &=~ $(\inf, \sup)$ & \\
        $\sem{\boxtimes}^\pparam_\bnd (\overrightarrow{\gsw, \inf_i, \sup_i, \eps_i})$ &=~ $(\gsw, \inf, \sup, f'(\eps_1, \eps_2))$ & $f'(\eps_1, \eps_2) \le q/t$\\
        $\sem{\boxdot}^\pparam_\bnd (\gsw, \inf_1, \sup_1, \eps_1), (\lwe, \inf_2, \sup_2, \eps_2)$ &=~ $(\rlwe, \inf, \sup, g(\eps_1, \eps_2))$ & $g(\eps_1, \eps_2) \le q/t$\\
        $\sem{\boxdot}^\pparam_\bnd (\gsw, \inf_1, \sup_1, \eps_1), (\rlwe, \inf_2, \sup_2, \eps_2)$ &=~ $(\rlwe, \inf, \sup, g'(\eps_1, \eps_2))$ & $g'(\eps_1, \eps_2) \le q/t$\\

    %%    \hline
    %%\multicolumn{5}{c} 
        && \qquad where \\
            && {$\inf := \min \{{\inf}_1 * {\inf}_2, {\inf}_1 * {\sup}_2 , {\sup}_1 * {\inf}_2, {\sup}_1 * {\sup}_2 \}$ }\\
            && {$\sup := \max \{{\inf}_1 * {\inf}_2, {\inf}_1 * {\sup}_2 , {\sup}_1 * {\inf}_2, {\sup}_1 * {\sup}_2 \}$ }\\
    \hline
    $\sem{\times}^\pparam_\bnd (n,(\id, \inf, \sup, \eps))$ &=~ $(\id, \inf',  \sup',  |n * \eps|)$ & $\inf' = \min \{n*\inf, n*\sup \}$ and $\sup' = \max \{n*\inf, n*\sup \}$ \\
    && if $-t/2 \le  \inf' \le  \sup' < t/2$ and $|n * \eps| \le q/t$\\
    \hline
    $\sem{\code{pbs}}^\pparam_\bnd (\gsw, \inf_0, \sup_0, \eps_0) (\lwe, \inf, \sup, \eps)$ &=~ $(\lwe, \inf, \sup, \eps_b )$ & if $ \max\{ \log_2|\inf|, \log_2 |\sup|\} \le (\log_2(t) -1)$\\
    && if $-t/2 \le  \inf_0 \le  \sup_0 < t/2$ \\
        %%%
    \bottomrule
    \end{tabular}
    \label{tab:tfhe_ops-app}
    \end{small}

\caption{\ilatfhe: \ila model for TFHE scheme illustrating key homomorphic operations.
Note that input operand has the type $(id_i, \inf_i, \sup_i, \eps_i)$ (for ciphertext)}
\label{tab:tfhe_operations-app}
\end{table*}

\subsubsection{Instantiation of \ila Functional Correctness} 
The commutativity and downwards closure axioms are similar to that of \ilabgv.
Since \ilatfhe is a valid \ila model, we thus get Theorem~\ref{thm:msg_equiv_commands} for free.

\section{\ilabgv modswitch Typing Rule}\label{app:ilabgvtr}
As yet another example, we show the typing rule for \code{modswitch} operation using the bounds computation from Table~\ref{tab:operations-bgv}.
Recall that modulus switching reduces both the modulus level and noise on a ciphertext.
Rule \rulename{T-MS} states that modulus switching reduces the noise by $\frac{q_{\om -1}}{q_{\om}}$ with some rounding error $B_r$. Additionally, the level of the ciphertext is lowered from $\om$ to $\om -1$. Notice that the resulting noise $\eps'$ is always lower than $\eps$ due to decreasing chain property of modulus levels.

\begin{mathpar}
 \inferrule[T-MS] { \jdg {\Ga} {e} {\cimod{\eps}{\om}{\inf}{\sup} } \\ 0 \le \om - 1 < L \\ \eps' = \frac{q_{\om-1}}{q_{\om}} \eps + B_r \le l_{\om-1}} 
 { \jdg {\Ga} {\modswitch{e}} {\cimod{\eps'}{\om-1}{\inf}{\sup}} } 
\end{mathpar}

\section{\texorpdfstring{\ila}{} Extensions}

\begin{figure}
\centering
\begin{syntax}
\categoryFromSet[Sorts]{s}{\{\MSort, \PSort, \CSort\}}
    \categoryFromSet[Bounds]{\alpha_s}{B_s}
    \category[Types]{\tau}
    \alternative{\Cipher\ \alpha_\CSort}
    \alternative{\Plain\ \alpha_\PSort}
    \alternative{\msg\ \alpha_\MSort}\\
    \category[Contexts]{\Gamma}
    \alternative{\cdot}
    \alternative{\Gamma, x:\tau}
\end{syntax}
\caption{\ila Extensions}\label{Extensions}
\end{figure}

\begin{figure}
\begin{mathpar}
\judgment{\pparam; \Gamma \vdash e : \tau} \and
\inferrule{x : \tau \in \Gamma}{\pparam; \Gamma \vdash x : \tau} \and
\inferrule{\sort(v) = s}{\pparam; \Gamma \vdash v : s\ |v|^\pparam_s} \and
\inferrule{\pparam; \Gamma \vdash e_i : s \, \alpha_i}{\pparam; \Gamma \vdash [e_1,\dots,e_n] : \vecs{s}\, (\alpha_1, \dots, \alpha_n)} \and
\inferrule{\op : \overrightarrow{s_i} \to s \and 
(\forall i, \pparam; \Gamma \vdash e_i : \tau_i \and \sort(\tau_i) = s_i) \and 
    \sem{\op}_\bnd^\pparam(\overrightarrow{|\tau_i|}) = \alpha
}{\pparam; \Gamma \vdash 
\op(\overrightarrow{e_i}) : s\ \alpha}
\end{mathpar}
\caption{\ila Typing: Expression including vectors }\label{app:ilaexptr}
\end{figure}

\end{document}